\documentclass[11pt]{article}
\usepackage[utf8]{inputenc}
\usepackage{geometry}
\usepackage{graphicx}
\usepackage{amsmath, amssymb}
\usepackage{hyperref}
\usepackage{setspace}
\usepackage{csquotes}
\usepackage[super,sort&compress]{natbib}
\title{QUIET: \textbf{Q}uantifying \textbf{U}nderutilized \textbf{I}nfluential \textbf{E}dges for \textbf{T}argeted Synchronization}

\author{Sovesh Mohapatra$^{1}$, Christoffer G. Alexandersen$^{1}$, Panagiotis Fotiadis$^{2}$, Max B. Kelz$^{3}$, \\ John A. Detre$^{4}$, 
Fabio Pasqualetti$^{5}$, Dani S. Bassett$^{1,4,\ast}$ \\[3pt]
{\small $^{1}$Department of Bioengineering, School of Engineering and Applied Science, University of } \\[-6pt]
{\small Pennsylvania, PA, USA}\\[-6pt]
{\small $^{2}$Department of Anesthesiology, University of Michigan, MI, USA} \\[-6pt]
{\small $^{3}$Department of Anesthesiology and Critical Care, Perelman
School of Medicine, University of} \\[-6pt]
{\small Pennsylvania, PA, USA}\\[-6pt]
{\small $^{4}$Department of Neurology, Perelman School of Medicine, University of Pennsylvania, PA, USA} \\[-6pt]
{\small $^{5}$Department of Electrical Engineering and Computer Science, University of California Irvine, CA, USA} \\[-6pt]
{\small $^{\ast}$Corresponding author: \texttt{dsb@seas.upenn.edu}}
}

\date{}
\begin{document}

\maketitle
\pagebreak
\section*{Abstract}

Network control theory can be used to model intrinsic and extrinsic strategies to steer neural dynamics. Standard approaches are node-centric, structural, and focused on achieving desired instantaneous states. Here, we develop an edge-centric approach which incorporates both structure and function to achieve extended patterns of neural dynamics characterized by desired synchronization states. Our method, Quantifying Underutilized Influential Edges for Targeted Synchronization (QUIET), is an edge-centric framework that integrates structural controllability of individual white matter connections and mutual information between pairwise functional timeseries to identify energy-efficient synchronization pathways. QUIET identifies \textit{quiet highways}, edges that are structurally influential but functionally underutilized, to optimize regional synchronization. We validated QUIET across 75 synthetic configurations, where QUIET-ranked edge sets significantly outperformed random selection in 93\% of cases ($p<0.01$). The framework, tested on Human Connectome Project participants, revealed that the control energy required for synchronization of the salience network correlates with fluid intelligence. QUIET, applied to healthy adults undergoing dexmedetomidine-induced unresponsiveness, showed that the frontoparietal and default-mode networks exhibited the largest control energy required for synchronization in both awake and sedated states. QUIET is released as a stand-alone software to be used to study theoretically-defined synchronization pathways, which in turn could inform testable hypotheses in perturbative studies.

\pagebreak

\section{Introduction}

Human brains navigate a complex energy landscape, steering neural activity through a relatively fixed structural connectome to support diverse cognitive functions \cite{gu2018energy, bassett2017network, bullmore2009complex}. The white-matter scaffold constrains signal propagation and shapes the range of accessible functional states \cite{honey2009predicting, deco2011emerging}. Transitions between these functional states require coordinated control energy to overcome topological constraints \cite{tang2017developmental, sun2023neonatal, bassignana2022aging, singleton2022receptor}. Network control theory (NCT) provides a framework for quantifying these costly coordinated dynamics \cite{pasqualetti2014controllability, gu2015controllability}. This approach treats the brain as a dynamical system in which control energy is applied to specific nodes to steer the network to a target state, thereby providing a quantification of the energetic cost of brain state transitions, with relevance for cognitive function, development, and pathology \cite{cornelius2013realistic, yan2017network, tang2018control, hahn2023genetic}.

Despite these advances, standard NCT is fundamentally node-centric, applying control inputs to brain regions and implicitly assuming that each structural connection transmits a control signal at full capacity, irrespective of how much endogenous activity it already carries \cite{avena2018communication}. This assumption may diverge from the biology particularly in rich-club hubs, which are metabolically expensive, densely interconnected regions thought to support global information integration \cite{vandenheuvel2011rich, bullmore2012economy}. While structurally central, these regions possess dense intrinsic functional coupling that may induce signal saturation, limiting the efficacy of external control inputs in ways not captured by standard NCT models \cite{karrer2020practical}. Moreover, applying control energy to a node propagates the signal along all efferent white-matter tracts, without accounting for the potential existence of separate task-relevant pathways and quiescent pathways \cite{sun2024edgecentric}. These limitations motivate an edge-centric perspective that also reformulates the nature of the control intervention itself. Rather than an exogenous input acting on nodes of a linear system, as in classical NCT \cite{pasqualetti2014controllability, gu2015controllability}, control is enacted here as a targeted modulation of the coupling weights of specific structural connections, enabling more precise and energy-efficient network steering.

The energetic perspective on synchronization has direct clinical relevance. In pathological states such as schizophrenia, the brain fails to navigate the energy landscape efficiently, requiring significantly higher control energy to transition between cognitive states \cite{braun2021brain, wang2022alterations, tang2023altered}. Under pharmacological interventions such as deep sedation, by contrast, the disruption is not one of elevated energy cost but of topological reorganization: anesthetics fundamentally alter the roles of network hubs and disrupt global integration, trapping the system in local energetic minima that differ qualitatively from those seen in psychiatric illness \cite{boveroux2010breakdown, luppi2019consciousness}. While standard node-centric control metrics have successfully characterized energy demands in each of these regimes \cite{braun2021brain, broeders2024energy, zoller2021psychosis}, they treat regional controllability as a static property and propagate signals indiscriminately along all efferent connections, providing limited guidance on which specific white-matter pathways mediate these distinct patterns of energy reorganization \cite{karrer2020practical}. In parallel, information-theoretic approaches can identify which regions act as hubs through which a disproportionate share of network information is routed, but they lack the generative power to predict how the system will respond to perturbation \cite{palmigiano2017flexible, timme2018tutorial}. These complementary approaches can be pursued independently, but no existing framework integrates them at the edge level to jointly identify energy-efficient intervention targets in large-scale brain networks.

\begin{figure}[htb]
    \centering
    \includegraphics[width=1\textwidth]{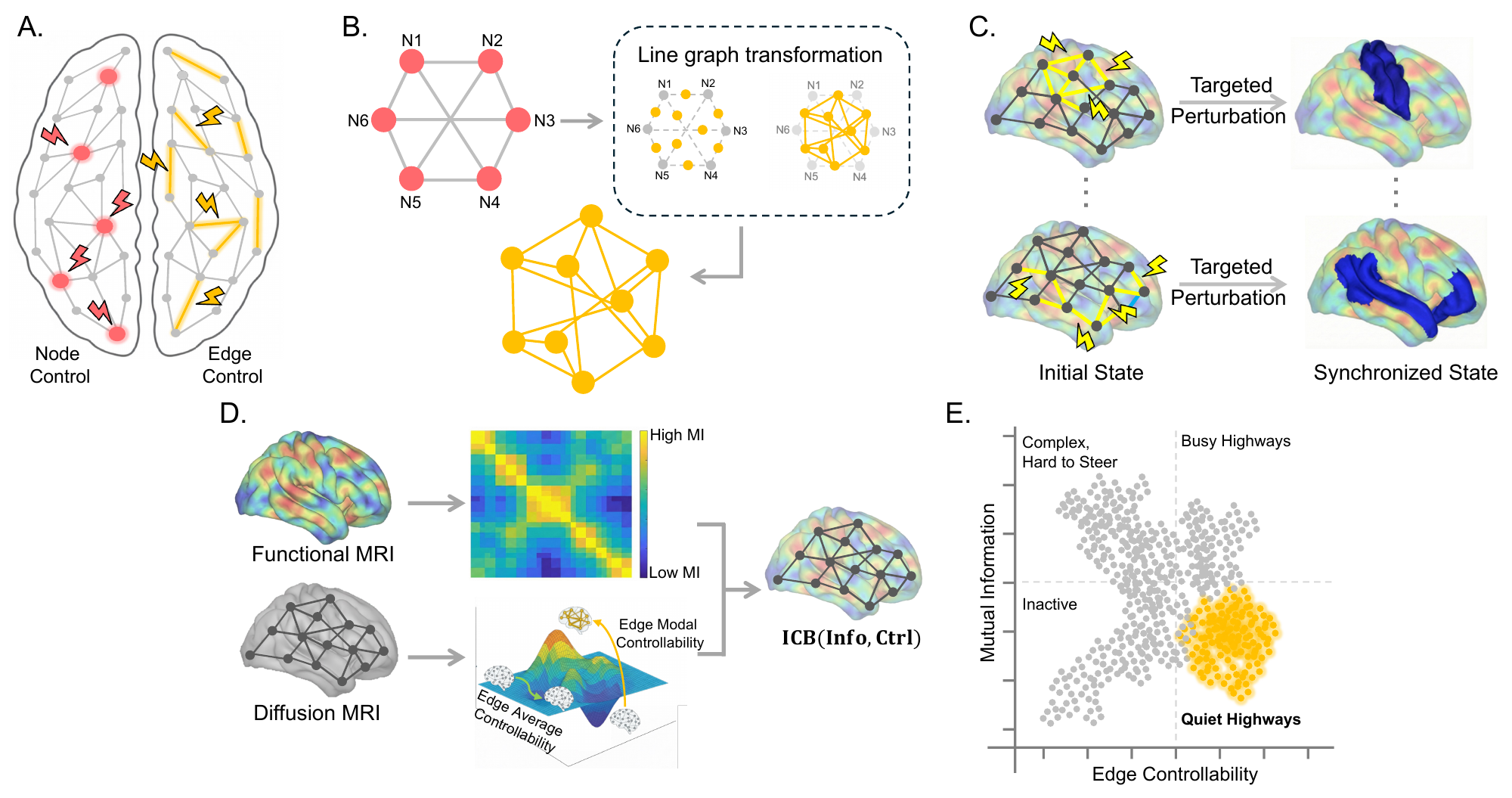}
    \caption{\textbf{QUIET reframes network control as an edge-level problem.}
    \textbf{A.} Node-centric control (left) applies perturbations to brain regions, propagating signals indiscriminately along all efferent connections; edge-centric control (right) targets specific white-matter tracts, enabling pathway-selective perturbation.
    \textbf{B.} The structural connectome is transformed into a line graph (or edge-to-vertex dual) in which each edge becomes a node and shared endpoints define adjacency. Since the edges of the original connectome are now the nodes of the line graph, standard node-controllability metrics yield edge-level values directly. \textbf{C.} QUIET identifies a targeted subset of edges (yellow lightning bolts) whose perturbation drives a designated cortical region from its initial desynchronized state to a final synchronized state (blue).
    \textbf{D.} Two complementary data streams feed the QUIET score: resting-state functional MRI yields an edge-wise mutual information (MI) matrix that quantifies functional coupling, while diffusion MRI yields a structural connectome from which edge average controllability (eAC) and edge modal controllability (eMC) are computed via the line graph.
    \textbf{E.} Each edge is plotted by its controllability ($x$-axis) and mutual information ($y$-axis). QUIET prioritizes \textit{quiet highways} (mustard yellow), edges that are structurally influential yet functionally underutilized, over busy highways (high controllability, high MI), complex, hard-to-steer edges (low controllability, high MI), and inactive edges (low controllability, low MI).}
    \label{Fig1}
\end{figure}

Here, we introduce the Quantifying Underutilized Influential Edges for Targeted Synchronization (QUIET), an edge-centric framework that integrates structural control theory and information-theoretic measures of functional connectivity (Fig. \ref{Fig1}A,D). Unlike node-centric approaches, QUIET operates on edges and ranks each white-matter connection by its structural controllability relative to its functional redundancy, as measured by mutual information (Fig. \ref{Fig1}B,E). This formulation identifies pathways that are structurally influential yet functionally underutilized and therefore, most amenable to energy-efficient perturbation (Fig. \ref{Fig1}C). We validate QUIET in three ways. First, we demonstrate in five different network topologies and three spatial scales that QUIET-ranked edge sets significantly reduce the control energy required for synchronization relative to randomly ranked edges. Next, we apply QUIET to 100 unrelated participants from the Human Connectome Project, revealing network-specific biomarkers of cognition and emotion. Finally, we test QUIET in an independent anesthetic dataset and show that sedation increases the QUIET-derived control energy required to synchronize functional brain networks. To facilitate the reproducible use of these techniques, we provide QUIET as a stand-alone, open-source software application.

\begin{figure}[htb]
    \centering
    \includegraphics[width=1\textwidth]{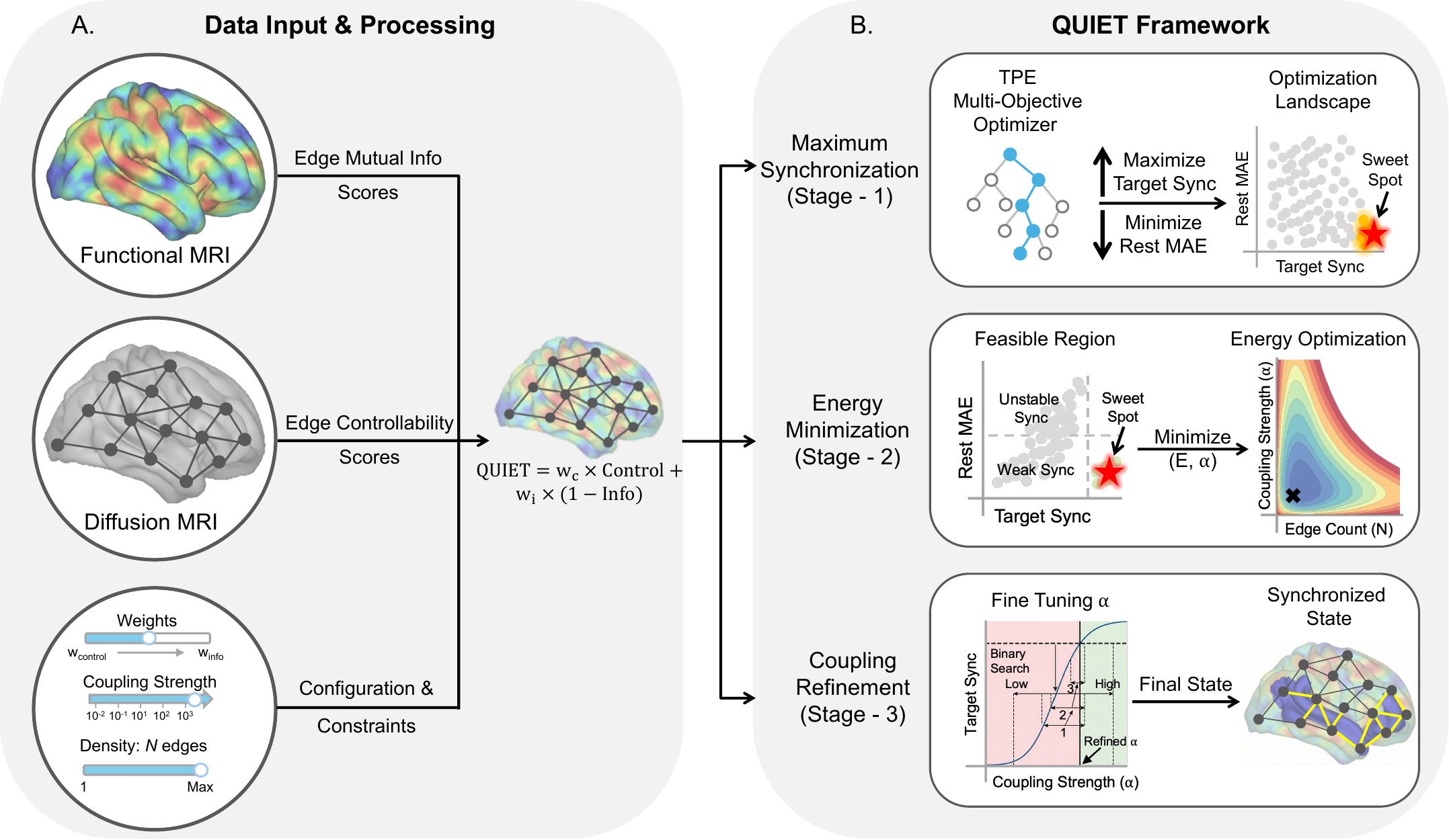}
    \caption{\textbf{QUIET integrates structural controllability and mutual information through a three-stage optimization pipeline.}
    \textbf{A.} Data input and processing. Resting-state functional MRI provides edge-wise mutual information (MI) scores that quantify functional coupling, while diffusion MRI provides edge-level controllability scores derived from the line graph of the structural connectome. These two feature sets are combined into the composite QUIET score. User-specified configuration parameters include the ranking weights, coupling strength, and the number of perturbed edges.
    \textbf{B.} Three-stage optimization. In Stage~1 (top), a tree-structured Parzen estimator (TPE) jointly maximizes target synchronization and minimizes off-target disruption (rest MAE); the Pareto front is evaluated and the sweet spot (star) closest to the ideal corner is selected. In Stage~2 (middle), the synchronization requirement is fixed at the sweet-spot target and the optimizer re-explores ranking weights and coupling strengths to minimize total control energy. In Stage~3 (bottom), the edge set is held fixed and a binary search over coupling strength $\alpha$ converges on the lowest value that still satisfies the synchronization constraint, yielding the final synchronized state.}
    \label{Fig2}
\end{figure}

\section{Results}

\subsection{QUIET integrates edge controllability and mutual information}

The QUIET framework reframes network control as an edge-ranking problem. QUIET operates on the line graph (or edge-to-vertex dual) of the structural connectome, in which each white-matter tract becomes a node and two tracts that share an endpoint become neighbors \cite{sun2024edgecentric}. For example, edges $(A, B)$ and $(B, C)$ in the original connectome share endpoint $B$ and therefore become adjacent nodes in the line graph, while edges with no common endpoint remain non-adjacent. This transformation enables the direct computation of edge-level average and modal controllability, which quantify how effectively each individual connection can steer the network toward easy-to-reach or difficult-to-reach states, respectively \cite{gu2015controllability, pasqualetti2014controllability} (Fig. \ref{Fig2}A and \nameref{Methods}). 

To complement these structural measures, QUIET quantifies the functional coupling between the two regions connected via each edge by computing pairwise mutual information from their resting-state time series using the Kraskov–Stögbauer–Grassberger (KSG) $k$-nearest-neighbor estimator \cite{kraskov2004estimating} (Fig. \ref{Fig2}A and \nameref{Methods}). Unlike linear correlation measures, the KSG estimator captures nonlinear statistical dependencies between neural signals without assumptions about the underlying distribution. The structural and information-theoretic scores are then combined into a single score for each edge $(i,j)$, where $i$ and $j$ denote the two brain regions (nodes) it connects. For each edge, the control term averages the percentile ranks of edge average controllability (eAC) and edge modal controllability (eMC), while the information term computes $1 - \mathrm{rank}(\mathrm{MI})$, which inverts the mutual information score so that functionally underutilized edges receive higher rankings. The final $\mathrm{QUIET}_{ij}$ score is a weighted sum of these two terms:

\begin{equation}
    \mathrm{QUIET}_{ij} = w_{\mathrm{ctrl}} \times \frac{\mathrm{rank}(\mathrm{eAC}_{ij}) + \mathrm{rank}(\mathrm{eMC}_{ij})}{2} + w_{\mathrm{info}} \times (1 - \mathrm{rank}(\mathrm{MI}_{ij})),
\end{equation}

\noindent where $w_{\mathrm{ctrl}}$ and $w_{\mathrm{info}}$ are not fixed hyperparameters but are instead discovered jointly with the coupling parameters by the optimizer in the first two stages of the optimization pipeline (Fig. \ref{Fig2}B and \nameref{Methods}). The ranking is regenerated at each optimization iteration, allowing the framework to continuously adapt the balance between structural influence and functional redundancy as it converges on the lowest-energy perturbation.

QUIET determines this perturbation through a three-stage optimization pipeline (Fig. \ref{Fig2}B and \nameref{Methods}). In Stage 1, a tree-structured Parzen estimator (TPE) jointly optimizes the ranking weights, edge count, and coupling strength to maximize target-region synchronization while minimizing off-target perturbation \cite{bergstra2011algorithms}. The resulting Pareto front is then evaluated, and the optimal synchronization target is identified by balancing the trade-off between maximum synchronization and minimal system disruption (Fig. \ref{Fig2}B and \nameref{Methods}). In Stage 2, the synchronization requirement is fixed at the optimal target found in Stage 1, and the optimizer re-explores ranking weights alongside coupling strength to minimize total control energy, defined as the product of the number of perturbed edges and their coupling strength (Fig. \ref{Fig2}B and \nameref{Methods}). In Stage 3, the coupling strength is refined via binary search to converge on the lowest-energy configuration that still satisfies synchronization constraints (Fig. \ref{Fig2}B and \nameref{Methods}). At each stage, network dynamics are simulated as coupled phase oscillators on a network, integrated with an adaptive Runge–Kutta method, and synchronization is quantified by the phase-locking value (PLV) computed after discarding initial transient timepoints \cite{kuramoto1984chemical, acebron2005kuramoto, lachaux1999measuring}. Detailed methodology is provided in the \nameref{Methods} section. Here, control energy quantifies a structural cost on the network's edges, specifically the total magnitude of coupling-weight modulation, rather than the input-norm energy of classical NCT.

\subsection{QUIET is robust across different network topologies}

\begin{figure}[htb]
    \centering
    \includegraphics[width=1\textwidth]{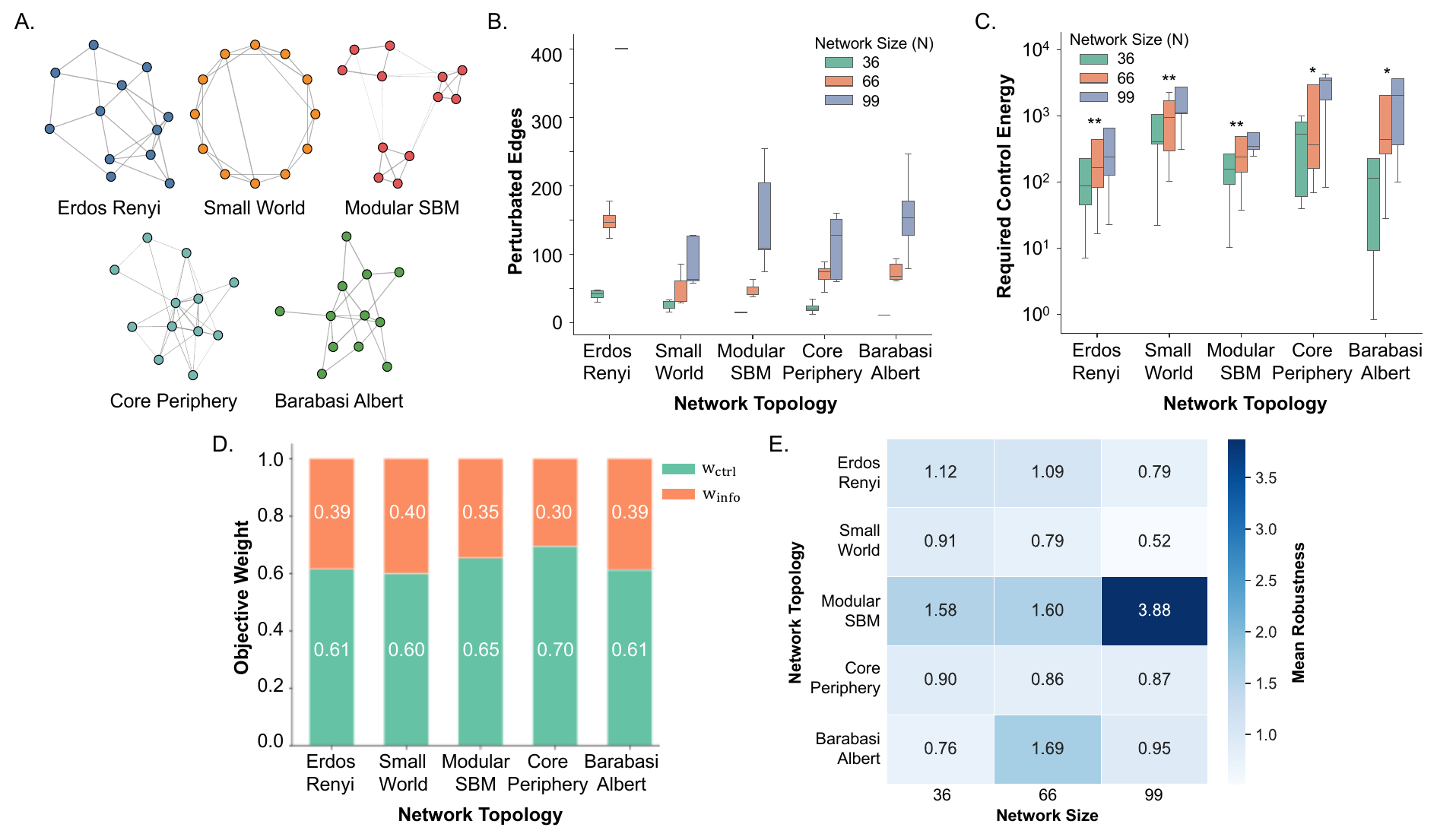}
    \caption{\textbf{QUIET generalizes across network topologies, scales, and coupling regimes.}
    \textbf{A.} Representative graphs of the five synthetic topologies.
    \textbf{B.} Number of perturbed edges selected by the framework for each topology at three network sizes ($N$ = 36, 66, 99) to reach the desired synchronization state.
    \textbf{C.} The control energy required across topologies and network sizes.
    \textbf{D.} Optimized ranking weights ($w_{\mathrm{ctrl}}$, green; $w_{\mathrm{info}}$, orange) averaged across scales and coupling strengths.
    \textbf{E.} Mean robustness across all topology-size combinations. Box plots show median, interquartile range, and $1.5\times$ IQR whiskers. $**$ all coupling-strength configurations significant ($p<0.01$); $*$ the majority of configurations significant.}
    \label{Fig3}
\end{figure}

The generalizability of QUIET was stress-tested across a broad range of network architectures: five synthetic topologies (Erdős-Rényi, small-world, modular stochastic block model, core-periphery, and Barabási-Albert) at three spatial scales ($N$ = 36, 66, 99 nodes) and five coupling strengths ($g$ = 0.01, 0.05, 0.1, 0.5, 1.0) (Fig. \ref{Fig3}A). In each configuration, a module (contiguous sub-network of nodes) was selected as the synchronization target, and QUIET was tasked with identifying the edge set whose perturbation would maximize phase coherence within that module while minimizing disruption to the rest of the network. For each of the 75 resulting configurations, we ran QUIET's three-stage optimization and compared the optimized edge set against 10,000 random permutations drawn from the same candidate pool.

The integration of structural controllability with mutual information is what drives QUIET's performance: the full ranking succeeded in 93\% of configurations, while controllability-only and mutual-information-only rankings, evaluated under identical optimization procedures, succeeded in only 28\% and 32\%, respectively (Supplementary Fig.~S1A-D). This near three-fold gain over the strongest single-component baseline establishes that the two channels are complementary rather than redundant: structural controllability identifies edges with mechanical leverage over the network, mutual information flags edges that carry distinguishable functional signals, and only their joint optimization resolves which connections both can and should be perturbed. The use of mutual information as the functional channel, rather than a simpler PLV-based alternative, was further supported by a direct comparison. When the entire pipeline was evaluated using baseline pairwise PLV (LeastPLV) instead of MI, it yielded statistically equivalent control energies (paired Wilcoxon $p = 0.40$) but significantly worse off-target stability ($p = 0.010$, Supplementary Fig.~S8), reflecting MI's capacity to read the full joint distribution of endpoint phases rather than the first circular moment alone. Against random selection, QUIET-ranked edge sets significantly outperformed the null in 70 of 75 configurations ($p<0.01$ against 10,000 permutations; Fig. \ref{Fig3}C), reaching a 100\% (15/15) success rate in Erdős-Rényi and modular stochastic block model topologies (Supplementary Fig.~S1C). The control energy required, here defined as the product of the number of perturbed edges and their coupling strength, spanned several orders of magnitude across topologies (Fig. \ref{Fig3}C). Erdős-Rényi networks required the largest number of perturbed edges but at very low per-edge coupling, whereas modular and small-world topologies achieved synchronization with fewer, more strongly coupled edges (Fig. \ref{Fig3}B). Small-world and core-periphery networks exhibited the highest mean energy requirements, while Barabási-Albert networks spanned the widest dynamic range across coupling regimes, with weak-coupling, large-$N$ configurations driving exceptional energy demands, consistent with the known difficulty of controlling scale-free architectures \cite{liu2011controllability, yan2012controlling}.

The optimizer adapted its ranking strategy to each topology, allowing QUIET to operate across architecturally distinct networks without manual retuning (Fig. \ref{Fig3}D). Across all five architectures, the control weight $w_{\mathrm{ctrl}}$ remained the dominant signal (0.60–0.70), confirming that structural controllability is the primary driver of edge selection. The relative emphasis on controllability scaled with the degree of mesoscale organization. Networks with strong block or hierarchical structure leaned most heavily on controllability, with core-periphery reaching $w_{\mathrm{ctrl}} = 0.70$ and the modular stochastic block model reaching $w_{\mathrm{ctrl}} = 0.65$. Networks lacking distinct community structure---Erdős–Rényi, Barabási–Albert, and small-world---ceded a substantial 39-40\% of the ranking weight to mutual information ($w_{\mathrm{info}} = 0.39$-$0.40$), such that identifying functionally redundant edges becomes as important as selecting structurally influential ones in the absence of strong structural cues. A single fixed weighting could not have produced these topology-appropriate selections, and the large gap between integrated and ablated rankings (Supplementary Fig.~S1A-D) is a direct consequence of this adaptive balance.

Robustness, quantified as the mean synchronization $z$-score relative to the null distribution, varied by network size and architectural complexity (Fig. \ref{Fig3}E). The modular stochastic block model at $N$ = 99 exhibited the highest mean robustness ($z_{\mathrm{sync}}$ = 3.88 averaged across coupling strengths, peaking at $z_{\mathrm{sync}}$ = 8.63 at $g$ = 0.05). This result indicates that the advantage of QUIET-ranked edges over random selection is most pronounced in large modular networks, where within-module edges differ sharply in their synchronization leverage and the candidate pool grows quadratically with module size, so random selection is increasingly unlikely to recover the few high-leverage edges that QUIET identifies. The mean $z$-score remained positive across every topology and every scale tested, establishing that the framework's superiority over chance-level edge selection holds uniformly across the entire 75-configuration sweep (Supplementary Figs.~S3-7).

\subsection{The control energy required for synchronization of functional networks captures behavioral variation}

We applied QUIET to the structural and functional connectomes of 100 unrelated participants from the Human Connectome Project (HCP), targeting the functional networks defined by the Schaefer 400-region parcellation: dorsal attention, limbic, frontoparietal, salience, and default-mode \cite{vanessen2013wuminn, schaefer2018local}. The sensorimotor and visual networks were excluded due to ceiling and floor effects in baseline synchronization, respectively (Supplementary Fig.~S2). For each subject, QUIET ranked the edge set for both the left- and right-hemisphere functional networks, and the resulting control energies required for synchronization were averaged across hemispheres to produce a single network-level estimate per subject.

\begin{figure}[htb]
    \centering
    \includegraphics[width=1\textwidth]{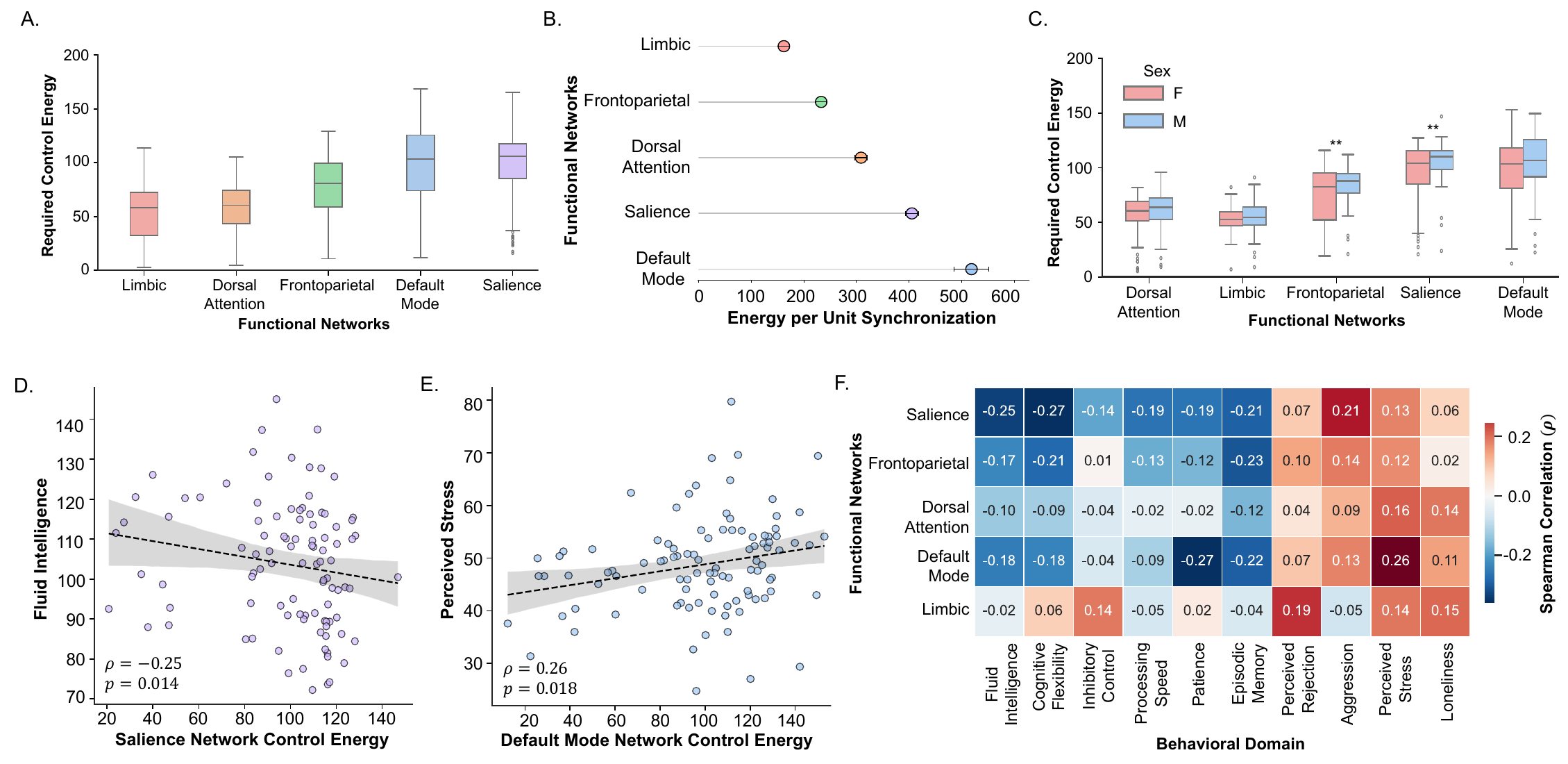}
    \caption{\textbf{The control energy required for synchronization of functional networks.}
    \textbf{A.} The control energy required for each of five functional networks in 100 HCP participants.
    \textbf{B.} Energy per unit of synchronization gain normalizes raw energy by the achieved synchronization improvement.
    \textbf{C.} Sex-stratified control energy. Females (F, pink) require significantly less energy than males (M, blue) to synchronize the frontoparietal and salience networks ($**$, $p<0.05$); no significant differences are observed in other networks.
    \textbf{D.} QUIET-derived control energy required for salience network synchronization is negatively correlated with fluid intelligence (Spearman $\rho=-0.25$, $p=0.014$). 
    \textbf{E.} QUIET-derived control energy required for default-mode network synchronization is positively correlated with perceived stress ($\rho=0.26$, $p=0.009$).
    \textbf{F.} Full energy-behavior landscape showing Spearman correlations between control energy of each functional network and ten cognitive and emotional measures. Box plots show median, interquartile range, and $1.5\times$ IQR whiskers. Shaded region denotes 95\% confidence interval. }
    \label{Fig4}
\end{figure}

Prior work has established that higher-order association networks such as the default-mode network are metabolically expensive at baseline \cite{raichle2015brain, bullmore2012economy} and sit at the top of a cortical hierarchy characterized by dense, distributed functional coupling \cite{vandenheuvel2011rich}. Hence, we hypothesized that the control energy required to synchronize these higher-order networks would exceed that required for more topologically constrained networks such as the limbic system. Consistent with our expectations, the salience and default-mode networks required the most energy to synchronize, whereas the limbic network required the least (Fig. \ref{Fig4}A). To account for differences in baseline synchronization, we computed the energy per unit of synchronization gain for each network, revealing that the limbic network was the most efficient to control, requiring approximately three-fold less energy per unit gain in synchronization than the default-mode network (Fig. \ref{Fig4}B).

Prior work demonstrates that sex differences in brain network organization are most pronounced in higher-order association networks rather than in unimodal sensory or limbic regions \cite{shanmugan2022sex}. Beyond these regional patterns, people assigned female at birth (AFAB) have been observed to exhibit greater local efficiency than people assigned male at birth (AMAB), with the effect more pronounced in smaller brains \cite{yan2011sex}. We therefore anticipated that AFAB people would require less control energy than AMAB people to synchronize the higher-order association networks. As we expected, AFAB people required significantly less control energy than AMAB people in the frontoparietal ($\Delta = -9.46$ arbitrary units (a.u.), $p=0.047$) and salience ($\Delta = -10.97$ a.u., $p=0.039$) networks; the dorsal attention, limbic, and default-mode networks showed no significant sex differences (Fig. \ref{Fig4}C). QUIET recovered this sex-dependent asymmetry from connectome inputs alone, with no demographic variables in the model. The recovered pattern follows directly from the greater local efficiency that AFAB people exhibit in higher-order association networks \cite{yan2011sex, shanmugan2022sex}, which predicts lower synchronization cost \cite{bullmore2012economy, cabral2011role}. Fluid intelligence was statistically indistinguishable between the two sexes studied (F: 105.88, M: 101.25; $t = 1.47$, $p = 0.144$; no intersex people are known to have participated in the study). This equivalence indicates that the sex-dependent energy asymmetry reflects how the network is organized rather than how well it performs, and supports the generalization of energy-behavior associations across these two sexes.

Beyond group-level differences, QUIET-derived control energy also captures continuous individual variation in cognition and emotions (Supplementary Table~S1). The salience network is widely recognized as a "switch" between default-mode and executive states, responsible for allocating attentional resources \cite{seeley2007dissociable, menon2010saliency, uddin2015salience}. We theorized that individuals with more efficiently synchronized salience networks would exhibit higher fluid intelligence. Consistent with this expectation, lower salience network control energy correlated with higher fluid intelligence ($\rho=-0.25$, $p=0.014$; Fig. \ref{Fig4}D). Similarly, because difficulty modulating default-mode activity is linked to maladaptive rumination \cite{raichle2015brain, whitfieldgabrieli2012default}, we expected that higher energy requirements to control this network would track with greater perceived stress. This association was indeed observed: higher default-mode control energy were associated with increased stress levels ($\rho=0.26$, $p=0.009$; Fig. \ref{Fig4}E).

These individual trait associations were part of a broader pattern revealed by the full energy–behavior landscape (Fig. \ref{Fig4}F). Cognitive and emotional measures were tested separately, and $p$-values were corrected for multiple comparisons using FDR within each behavioral domain per network. The salience network showed the strongest and most widespread negative correlations with cognitive measures: cognitive flexibility ($\rho=-0.27$, $q_{\mathrm{FDR}}=0.041$) and fluid intelligence ($\rho=-0.25$, $q_{\mathrm{FDR}}=0.041$) both survived FDR correction, whereas episodic memory showed a trend ($\rho=-0.21$, $p=0.032$, uncorrected), consistent with its established role as a domain-general hub for cognitive resource allocation \cite{uddin2011dynamic}. The salience network also showed a positive correlation with aggression ($\rho=0.21$, $p=0.033$, uncorrected), suggesting that individuals whose salience networks require more energy to synchronize exhibit greater trait aggression. The default-mode network was negatively correlated with patience ($\rho=-0.27$, $q_{\mathrm{FDR}}=0.046$) and episodic memory ($\rho=-0.22$, $p=0.031$, uncorrected), linking difficulty in modulating default-mode activity to both impulsive decision-making and poorer memory consolidation; in the emotional domain, perceived stress survived FDR correction ($\rho=0.26$, $q_{\mathrm{FDR}}=0.036$). The frontoparietal network exhibited negative correlations with episodic memory ($\rho=-0.23$, $p=0.024$, uncorrected) and cognitive flexibility ($\rho=-0.21$, $p=0.038$, uncorrected), consistent with its established role in executive control \cite{cole2013multitask}. Processing speed and inhibitory control showed trends in the expected direction for the salience network ($p=0.053$ and $p=0.166$, respectively, uncorrected) but did not reach significance, and no associations survived for the dorsal attention or limbic networks individually. This network-specific pattern of energy–behavior associations demonstrates that the control energy required for synchronization is not merely a structural property of the connectome but a behaviorally meaningful biomarker, capturing individual differences that span cognitive flexibility, attentional control, and emotional regulation.

\subsection{Dexmedetomidine sedation reshapes network-level control energy}

The sensitivity of the QUIET framework to pharmacological state changes was assessed in an independent cohort of participants undergoing dexmedetomidine-induced sedation \cite{fotiadis2025changes}. Each subject was scanned in both awake and sedated states, providing a within-subject paired design. The control energy was computed for the same five functional networks as in the HCP analysis: dorsal attention, limbic, frontoparietal, salience, and default-mode.

\begin{figure}[htb]
    \centering
    \includegraphics[width=1\textwidth]{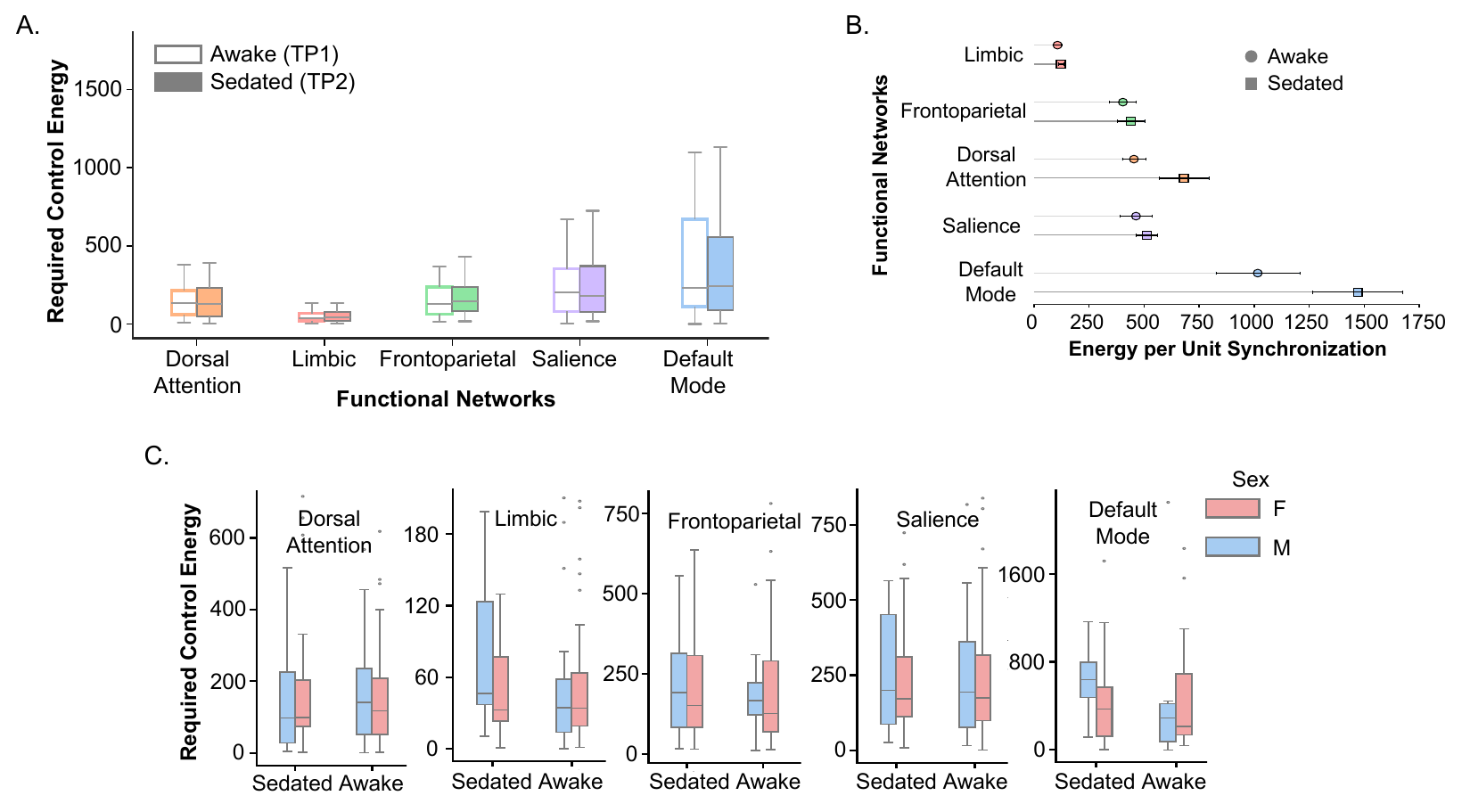}
    \caption{\textbf{Sedation tends to increase the energetic cost of network synchronization.}
    \textbf{A.} The control energy required for five functional networks in participants scanned while awake (TP1, open boxes) and under dexmedetomidine-induced sedation (TP2, filled boxes).
    \textbf{B.} Energy per unit of synchronization gain in awake (circle) and sedated (square) states.
    \textbf{C.} Sex-stratified control energy across awake and sedated states for each network. Females (F, pink) require less control energy than males (M, blue) in both conditions for synchronization of the limbic, salience, and default-mode networks; this pattern reaches significance for the limbic network under sedation ($p<0.05$). Box plots show median, interquartile range, and $1.5\times$ IQR whiskers.}
    \label{Fig5}
\end{figure}

Anesthetics disrupt long-range neural coordination and reduce both local and global network efficiency \cite{hashmi2017dexmedetomidine, luppi2019consciousness, demertzi2019human}. We therefore expected that sedated brains would require more control energy than awake brains to achieve the same level of network synchronization. Sedation indeed showed a trend toward increased control energy in most of the five networks (Fig. \ref{Fig5}A), with the frontoparietal and default-mode networks exhibiting the largest absolute increases, the limbic network showing a more modest increase, and the salience and dorsal attention networks showing slight decreases. To account for differences in achievable synchronization gain, we normalized the control energy by the increase in synchronization for each network in both states (Fig. \ref{Fig5}B). The limbic network was the least costly per unit gain in both conditions, whereas the default-mode network required approximately nine-fold more energy per unit of synchronization, replicating the network-level efficiency hierarchy observed in the HCP cohort. Sedation increased the per-unit cost across all networks, indicating that the sedated brain demands more energy to achieve the same degree of synchronization.

The sex-stratified analysis within the sedation cohort revealed a consistent trend in which AFAB people required less control energy than AMAB people in most networks and states (Fig. \ref{Fig5}C). This pattern in the limbic network during sedation was statistically significant ($\Delta = -29.4$ a.u., $p=0.048$). Although the remaining networks did not reach statistical significance individually, likely reflecting the limited cohort size ($n=14$), the directional consistency between networks aligns with the view that AFAB people exhibit more efficient controllability of higher-order association networks. Previous work has also established that AFAB people exhibit distinct dose-response profiles and faster emergence from general anesthesia, pointing to sex-dependent differences in anesthetic sensitivity at both pharmacokinetic \cite{buchanan2009patient} and pharmacodynamic \cite{wasilczuk2024hormonal} levels.
Together, these findings suggest that sex-linked differences in network controllability may be a structural feature of the connectome that persists across pharmacological states, though larger sedation cohorts will be needed to establish this relationship conclusively.

\subsection{Network synchronization software}

\begin{figure}[htb]
    \centering
    \includegraphics[width=1\textwidth]{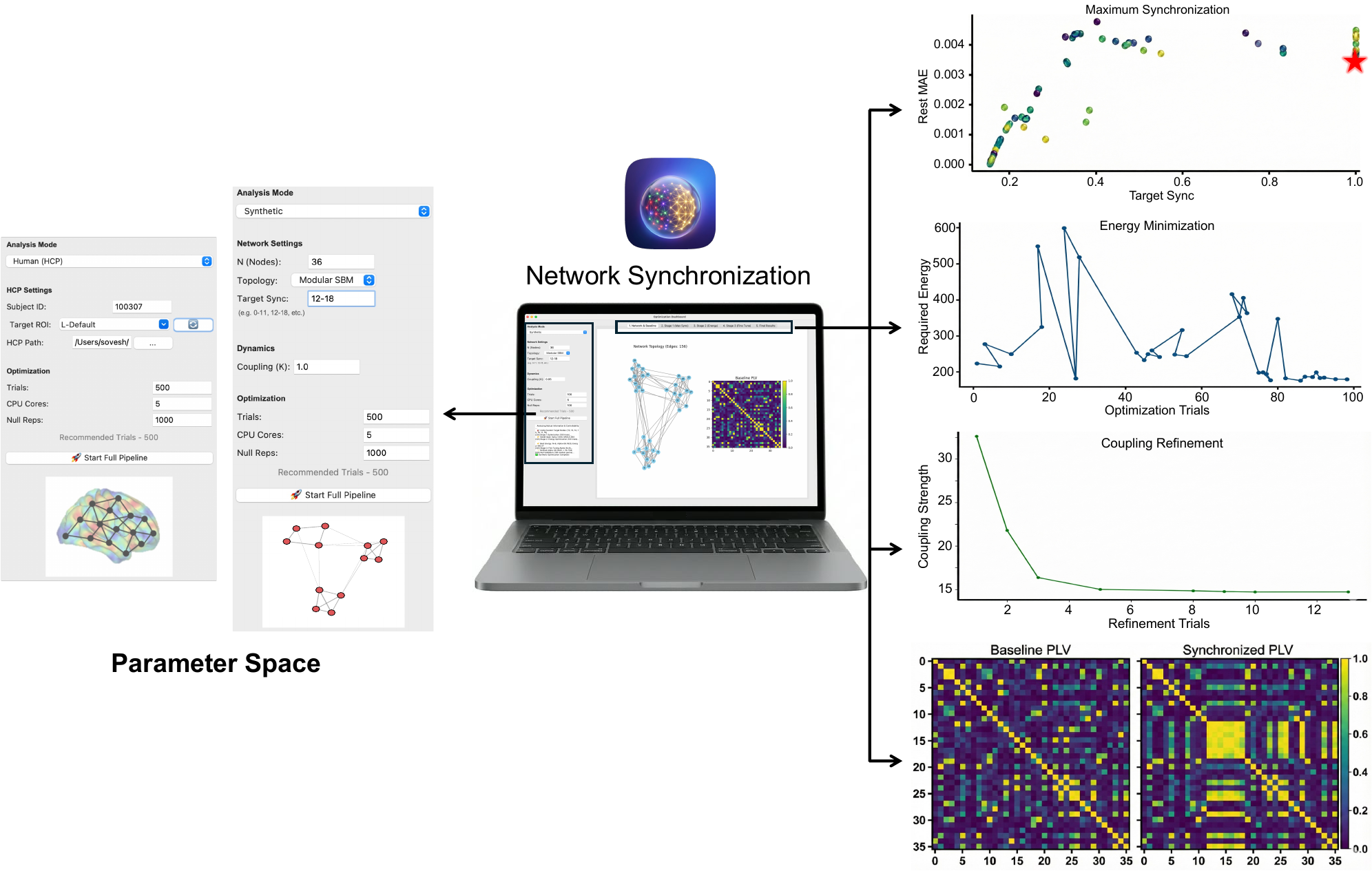}
    \caption{\textbf{Network Synchronization: open-source software for edge-centric control analysis.}
    The application provides two analysis modes. In the synthetic mode (left panel), users configure the network topology, size, target nodes, coupling dynamics, and optimization parameters. In the empirical mode (left panel), users specify a subject identifier, target functional network, and dataset path. Both modes execute the full three-stage QUIET pipeline and return interactive output visualizations (right). The maximum synchronization panel (top right) shows the Pareto front of target synchronization versus off-target disruption (rest MAE), with the sweet-spot optimization iteration (star) selected as the operating point. The energy minimization panel traces the control energy required across optimization iterations. The coupling refinement panel shows the binary-search convergence of coupling strength $\alpha$ to its minimum feasible value. The baseline and synchronized phase-locking value (PLV) matrices (bottom right) display the pairwise synchronization structure before and after the QUIET-optimized perturbation.}
    \label{Fig6}
\end{figure}

To facilitate reproducible use of the QUIET framework, we developed Network Synchronization, an open-source desktop application with a graphical user interface (Fig. \ref{Fig6}). The application supports two analysis modes: a synthetic mode, in which users specify network topology, size, coupling dynamics, and optimization parameters; and an empirical mode, in which users supply subject identifiers and select target functional networks from different datasets. In both modes, the software executes the full three-stage QUIET pipeline: maximum synchronization, energy minimization, and coupling refinement, and returns interactive visualizations of the synchronization landscape, energy convergence, and baseline-versus-optimized phase-locking matrices.

\section{Discussion}
We have introduced the Quantifying Underutilized Influential Edges for Targeted Synchronization (QUIET), an edge-centric framework that integrates structural controllability and information-theoretic measures of functional coupling to identify energy-efficient synchronization pathways in brain networks. 
By shifting the unit of analysis from nodes to edges, QUIET addresses a gap that arises when connection-level specificity is required: standard node-centric approaches do not distinguish which white-matter connections should carry a control signal from those that should remain quiescent, and for applications such as targeted neuromodulation this distinction may be critical. The resulting edge-level ranking enables targeted, low-energy simulated perturbations that generalize across network architectures, imaging cohorts, and pharmacological states.

Existing approaches to brain network control in the recent literature have largely operated at the node level, treating entire brain regions as units of perturbation and quantifying their controllability through average and modal controllability metrics \cite{gu2015controllability, pasqualetti2014controllability}, the control energy required to drive state transitions \cite{singleton2022receptor, klickstein2017energy}, and the identification of driver nodes that govern network steering \cite{liu2011controllability, benmessaoud2025lowdim}. Although these measures have provided important insights into developmental pathways \cite{tang2017developmental, sun2023neonatal, bassignana2022aging, singleton2022receptor, cui2020executive, jamalabadi2023parenthood}, psychiatric conditions \cite{braun2021brain, wang2022alterations, tang2023altered, hahn2023genetic, zoller2021psychosis, singleton2024alcohol, yuan2026ocd, parkes2021psychosis, jeganathan2018bipolar, niu2026depression}, and the relationship between white-matter architecture and stimulation spread \cite{stiso2019white, fang2022personalizing, wang2025eeg, khambhati2019stimulation}, they cannot prescribe which connections to modulate. Despite node-level delivery of stimulation modalities such as transcranial magnetic stimulation and deep brain stimulation, edge-level rankings can stratify patients by their predicted response to a given stimulation target and guide multi-site protocols that recruit specific connections via their endpoints \cite{medaglia2015cognitive, beynel2020structural, hahn2023ect, fox2012efficacy, ozdemir2020individualized}. Conversely, edge-centric analyses of functional connectivity have demonstrated that individual connections carry distinct, subject-specific information about brain states \cite{faskowitz2020edge}, but lack a generative model to predict the dynamical consequences of perturbation. QUIET bridges this gap by embedding both structural controllability and functional redundancy into a single, optimizable edge-level score.

QUIET's ability to identify edges for effective synchronization generalizes across dynamical systems displaying five different network topologies, from homogeneous Erdős–Rényi to heterogeneous Barabási–Albert networks. This breadth establishes that the framework is not tuned to a single connectivity regime: a control-dominant ranking applies uniformly across all topologies, with information weighting providing graded modulation that scales with each architecture's mesoscale organization \cite{liu2011controllability, yan2012controlling}. When applied to empirical connectomes from the Human Connectome Project, the framework revealed that the energy required to synchronize a functional network is not merely a topological property but a behaviorally informative biomarker. The salience network, whose control energy was associated with fluid intelligence and cognitive flexibility, occupies a privileged position in cognitive architecture as the switch between the default-mode and frontoparietal networks \cite{menon2010saliency, uddin2015salience}. The association between lower salience-network energy and better cognitive performance suggests that individuals whose salience network is more easily reconfigured can allocate attentional resources more flexibly, a hypothesis testable with longitudinal or intervention designs. At the opposite end of the energetic hierarchy, the limbic network's exceptional control efficiency aligns with the evolutionary architecture of paralimbic cortex. Paralimbic territories sit near the base of the cortical hierarchy \cite{mesulam1998sensation, margulies2016situating} and mediate the rapid affective and motivational signals that drive behavior. Their sparse, modular structural organization is consistent with a design principle in which evolutionarily conserved drivers of behavior are made energetically cheap to synchronize. This interpretation is testable: paralimbic edge rankings should remain stable across developmental cohorts in which higher-order networks are still maturing. Notably, the sensorimotor and visual networks were excluded from the primary analysis because they exhibited distinct ceiling and floor effects, respectively (Supplementary Fig.~S2). The sensorimotor network displayed the highest baseline phase-locking values, leaving negligible room for optimization-driven gains. This strong intrinsic coherence aligns with predictions from computational models of large-scale resting-state dynamics \cite{cabral2011role, honey2007network}. In contrast, the visual network's synchronization remained near baseline despite full optimization. This resistance to distributed phase coupling may reflect the distinct structural connectivity and local oscillatory properties of visual areas, which have been modeled previously using similar coupled-oscillator frameworks \cite{cabral2011role, honey2007network}. More broadly, both excluded networks sit at the sensory end of the cortical hierarchy and are characterized by the strongest structure-function coupling in the brain \cite{baum2020development, vazquez2019gradients}. This tight coupling may render their dynamics less adaptable to external perturbation. Given that their functional organization is already well-predicted by structural connectivity, there is limited flexibility for edge-level control signals to reconfigure synchronization. In contrast, the attentional and transmodal networks examined here exhibit comparatively weaker structure-function coupling and greater functional flexibility \cite{baum2020development, vazquez2019gradients}, which may be precisely what makes them tractable theoretical targets for edge-centric control.

The analysis of an independent anesthesia dataset extends these findings from trait-level variation to state-level modulation. The trend toward increased control energy under dexmedetomidine is consistent with the view that anesthetic agents do not simply attenuate neural activity but fundamentally alter the network's capacity for efficient information integration \cite{hashmi2017dexmedetomidine, luppi2019consciousness}. The fact that the default-mode network exhibited one of the largest increases in absolute energy aligns with converging evidence that suppression of default-mode connectivity is a hallmark of reduced responsiveness across pharmacological and pathological conditions \cite{boveroux2010breakdown, demertzi2019human, anticevic2012role}. The parallel increase in the frontoparietal network further supports the anesthetic disruption of higher-order association networks. The directionally consistent sex-dependent trends in this independent cohort, under a distinct pharmacological state and imaging protocol, strengthens the interpretation that these asymmetries reflect stable structural features of the connectome rather than artifacts of a single dataset or cognitive task. Together, these cross-cohort results suggest that QUIET-derived control energy required for synchronization captures both enduring individual traits and transient state changes.

The current framework has several constraints. QUIET models network dynamics with coupled phase oscillators \cite{kuramoto1984chemical, acebron2005kuramoto}, which capture synchronization but not the full biophysical complexity of neural populations \cite{neymotin2020hnn, tolley2024sbi, palmer2021maximally}. The structural connectomes were derived from diffusion tractography, which is subject to known fiber-tracking biases \cite{maier2017challenge}. The controllability metrics that initialize the ranking assume linear dynamics, an approximation that is particularly appropriate near operating points and that has been empirically validated for resting-state fMRI \cite{nozari2024macroscopic}; the resulting predictions may not generalize when the brain moves between distinct dynamical regimes \cite{karrer2020practical}. Despite these constraints, the QUIET framework opens several avenues for future work. The application of QUIET to neonatal connectomes using functional parcellations for the developing brain could reveal how synchronization cost evolves during ongoing processes of myelination and pruning \cite{MohSov_TReND_MICCAI2025}. An extension to task-evoked transitions would quantify the energy required to switch between cognitive states in clinical populations \cite{braun2021brain, cole2013multitask, vidaurre2017brain}. Finally, integration of QUIET edge rankings with targeted neuromodulation could enable principled stimulation target selection from individual connectomes \cite{medaglia2015cognitive, fox2012efficacy, beynel2020structural, ozdemir2020individualized}.

In summary, by unifying structural controllability with information-theoretic coupling at the level of individual white-matter connections, QUIET provides a principled framework for quantifying the theoretically predicted energetic cost of brain network synchronization. The accompanying open-source software makes these analyses accessible to the broader neuroscience community and generates testable predictions whose validation through targeted stimulation or perturbative experiments could ultimately inform individualized circuit-level intervention design.

\section{Methods}
\label{Methods}

\subsection{Data}

\subsubsection{Synthetic networks}

Synthetic data included five network topologies that span a range of structural properties: Erdős–Rényi ($p_{\mathrm{edge}} = 0.80$), Watts-Strogatz small-world networks ($k = 12$ nearest neighbors, rewiring probability $p = 0.2$), modular stochastic block models (3 modules; $p_{\mathrm{intra}} = 0.7$, $p_{\mathrm{inter}} = 0.05$), core-periphery networks (core size $= N/3$; $p_{\mathrm{core}} = 0.8$, $p_{\mathrm{core\text{-}periphery}} = 0.4$, $p_{\mathrm{periphery}} = 0.2$), and Barabási–Albert networks ($m = 12$ edges per new node). Each topology was constructed at three spatial scales ($N = 36$, $66$ and $99$ nodes) with a fixed random seed for reproducibility. Edge weights were drawn from topology-specific uniform distributions and each adjacency matrix was normalized to the unit mean node strength for cross-topology comparability. The natural frequencies of the phase oscillators were sampled from $\mathcal{N}(1.0,\, 0.25)$ and the initial phases from $\mathcal{U}(0,\, 2\pi)$. Five global coupling strengths ($g = 0.01$, 0.05, 0.1, 0.5 and 1.0) were applied to each configuration, yielding 75 unique topology-scale coupling combinations.

\subsubsection{Human Connectome Project}

A sample of 100 unrelated healthy participants (54\% female; mean age $29.1 \pm 3.7$ years; range 22-36 years) was drawn from the Human Connectome Project S1200 release \cite{vanessen2013wuminn, fotiadis2023myelination}. Participants were scanned on a customized Siemens Connectome Skyra 3\,T scanner (32-channel head coil) and underwent T1-weighted imaging (MEMPRAGE; 0.7\,mm isotropic; TR/TE $= 2{,}400/2.14$\,ms), resting-state fMRI (gradient-echo EPI; four runs $\times$ 1,200 volumes; 2\,mm isotropic; TR/TE $= 720/33.1$\,ms) and high angular resolution diffusion imaging (1.25\,mm isotropic; TR/TE $= 5{,}520/89.5$\,ms; $b_{\max} = 3{,}000$\,s/mm$^{2}$; 270 directions). Functional networks were defined using the Schaefer 400-region parcellation mapped to the Yeo 7-network atlas \cite{schaefer2018local, yeo2011organization}. Informed consent in writing was obtained from all participants in the Human Connectome Project, and procedures were approved by the Institutional Review Board of Washington University.

\subsubsection{Anesthesia cohort}

Fourteen healthy individuals (mean age $28.3 \pm 5.5$ years; range 22-39 years; 10 females) were scanned at the Hospital of the University of Pennsylvania on a Siemens Prisma 3\,T scanner (64-channel head/neck coil) \cite{fotiadis2025changes}. Imaging included T1-weighted (0.8\,mm isotropic; TR/TE $= 2{,}400/2.24$\,ms), diffusion-weighted (3\,mm isotropic; TR/TE $= 2{,}500/65$\,ms; 61 directions; multiband factor 2) and resting-state functional (700 volumes; 2\,mm isotropic; TR/TE $= 720/37$\,ms) sequences. Baseline scans were acquired while awake; dexmedetomidine was then administered as a bolus ($1\,\mu$g/kg over 10\,min) followed by continuous infusion ($1\,\mu$g/kg/h). Loss of responsiveness was confirmed by spontaneous eye closure and non-response to verbal commands, after which imaging was re-acquired. Informed consent was obtained from all individuals, and procedures were approved by the Institutional Review Board of the University of Pennsylvania.

\subsection{Processing pipelines}

\subsubsection{Human Connectome Project}

Structural connectomes were constructed using an atlas-based approach. Raw diffusion-weighted images were minimally pre-processed using the HCP consortium pipelines, which applied $b_0$ intensity normalization, EPI distortion correction, eddy-current and motion correction, gradient-nonlinearity correction, and registration to the subject's native T1-weighted scan \cite{glasser2013minimal}. Multi-shell, multi-tissue constrained spherical deconvolution was performed in MRtrix3 \cite{tournier2019mrtrix3} to estimate fibre orientation distributions, followed by anatomically constrained probabilistic tractography using the second-order integration over fibre orientation distributions (iFOD2) algorithm \cite{smith2012anatomically}. An initial whole-brain tractogram of ten million streamlines was generated per subject and refined using the SIFT2 reweighting scheme to reduce known tractography biases \cite{smith2015sift2, fotiadis2023myelination}. The refined tractogram was mapped to the Schaefer 400-region parcellation \cite{schaefer2018local} registered to native space, yielding a symmetric weighted structural connectivity matrix per subject; edge weights were defined as the SIFT2-weighted streamline sum divided by the summed grey-matter volumes of the connected regions \cite{fotiadis2023myelination}.

Resting-state fMRI data were pre-processed using the HCP consortium pipelines, which corrected for gradient distortion, subject motion, and EPI distortion, applied intensity normalization, and registered functional volumes to MNI space \cite{glasser2013minimal}. Cross-subject alignment was performed using areal-feature-based multimodal surface matching (MSMAll) \cite{glasser2016multi}, and structured independent component analysis with hierarchical fusion of classifiers (sICA+FIX) was applied for further denoising \cite{salimi2014automatic}. Run-level time series were demeaned and variance-normalized, then averaged across the four resting-state runs (1,200 volumes each) to produce a single representative time series per subject. A detailed description of the full processing pipelines can be found in Fotiadis \emph{et al.} (2023) \cite{fotiadis2023myelination}.

\subsubsection{Anesthetic cohort}

Diffusion-weighted images were denoised and corrected for eddy-current distortions and subject motion in MRtrix3 \cite{tournier2019mrtrix3}, followed by $B_1$ field inhomogeneity correction. Response functions were estimated using the \texttt{dhollander} algorithm, and fibre orientation distributions were obtained via multi-shell, multi-tissue constrained spherical deconvolution. An anatomically constrained whole-brain tractogram of ten million streamlines was generated and refined with the SIFT2 reweighting scheme \cite{smith2015sift2}. The refined tractogram was mapped to a 132-region combined Harvard-Oxford and AAL atlas (91 cortical, 15 subcortical, and 26 cerebellar regions), yielding a symmetric weighted structural connectivity matrix per subject; edge weights were defined as the SIFT2-weighted streamline sum divided by the summed grey-matter volumes of the connected regions \cite{fotiadis2025changes}.

Functional connectivity was derived from pseudo-continuous arterial spin labeling (pcASL) data (60 volumes; 3.75\,mm isotropic; TR/TE $= 4{,}000/10.3$\,ms). ASL images were motion-corrected, co-registered to the subject's T1-weighted anatomical scan, and nonlinearly normalized to MNI space using FSL's \texttt{fnirt} \cite{fotiadis2025changes}. Mean regional time series were extracted for each of the 132 atlas regions, and functional connectivity was quantified as the average coherence across all frequencies above 0.01\,Hz, computed via cross-power spectral density and normalized with a Fisher $r$-to-$z$ transform \cite{fotiadis2025changes}.

\subsection{Network of coupled phase oscillators}

Network dynamics were modeled with the coupled phase oscillator system composed of Kuramoto oscillators \cite{kuramoto1984chemical, acebron2005kuramoto}. Each node $i$ in a network of $N$ oscillators is assigned a natural frequency $\omega_i$ and a time-varying phase $\theta_i(t)$. The phase evolves according to:

\begin{equation}
    \frac{d\theta_i}{dt} = \omega_i + g \sum_{j=1}^{N} A_{ij} \sin\!\bigl(\theta_j - \theta_i\bigr), \qquad i = 1,\dots,N \qquad ,
\end{equation}

\noindent where $g$ is the global coupling strength and $A_{ij}$ denotes the elements of the weighted adjacency matrix $\mathbf{A} \in \mathbb{R}^{N \times N}$. For sparse networks (edge density $<0.4$), an equivalent incidence-matrix formulation was used to accelerate computation: given the oriented incidence matrix $\mathbf{B} \in \mathbb{R}^{N \times M}$ and edge-weight vector $\mathbf{w} \in \mathbb{R}^{M}$, the dynamics is equivalent to:

\begin{equation}
    \frac{d\boldsymbol{\theta}}{dt} = \boldsymbol{\omega} + g\,\mathbf{B}\bigl(\mathbf{w} \odot \sin(\mathbf{B}^{\!\top}\boldsymbol{\theta})\bigr) ~,
\end{equation}

\noindent where $\odot$ denotes element-wise multiplication. Both formulations were compiled with Numba for performance. The system was integrated with an adaptive Runge-Kutta method (RK4(5), Dormand-Prince), with relative and absolute tolerances of $10^{-8}$ and $10^{-10}$, respectively. Each simulation spanned $t \in [0, 40]$ arbitrary time units, evaluated at 800 equally spaced points.

Synchronization was quantified by the phase-locking value (PLV) \cite{lachaux1999measuring}. The first 20\% of simulated time points ($t < 8$) was discarded as transient, and PLV was computed over the remaining stable window. For each connected edge $(i,j)$, the edgewise PLV was defined as:

\begin{equation}
    \mathrm{PLV}_{ij} = \biggl\lvert \frac{1}{T} \sum_{t=1}^{T} e^{\,i\bigl(\theta_i(t) - \theta_j(t)\bigr)} \biggr\rvert ~,
\end{equation}

\noindent where $T$ is the number of post-transient time samples. PLV ranges from 0 (no phase coupling) to 1 (perfect synchronization). Target-region synchronization was evaluated by averaging PLV across all edges within the designated network.

\subsection{Empirical phase oscillator calibration}

The network of phase oscillators was optimized for the human datasets through a three-step calibration procedure. Resting-state BOLD time series were bandpass-filtered between 0.04 and 0.07\,Hz (fourth-order Butterworth, zero-phase), and the instantaneous phase was extracted via the Hilbert transform. The dominant natural frequency $\omega_i$ for each node was estimated from the peak of the within-band power spectrum and converted to angular frequency ($\omega_i = 2\pi f_i$). The structural adjacency matrix was globally scaled by a factor of $(1 + G_{\mathrm{glob}})$ with $G_{\mathrm{glob}} = 0.01$, and spatially correlated noise was injected at each integration step via the Cholesky factor of a structure-informed covariance matrix $\boldsymbol{\Sigma} = \alpha \mathbf{I} + \beta \mathbf{A}_{\mathrm{PSD}}$ ($\alpha = 0.3$, $\beta = 0.15$), where $\mathbf{I}$ is the identity matrix and $\mathbf{A}_{\mathrm{PSD}}$ is the symmetrized structural adjacency matrix shifted to positive semi-definiteness by subtracting its minimum eigenvalue when negative. The global coupling strength $g$ and noise amplitude $\sigma$ were calibrated per functional network by simulating the network of coupled phase oscillators over a grid of $g \in [0.10, 0.40]$ and $\sigma \in \{0.00, 0.05, 0.10, 0.15, 0.20\}$ and selecting the $(g, \sigma)$ pair that maximized the Pearson correlation between simulated and empirical PLV matrices. The effective parameters for the full 400-node simulation were set to the median of the per-network best values.

\subsection{Edge-wise mutual information}

Edge-level functional coupling was quantified using the Kraskov-Stögbauer-Grassberger (KSG) mutual information estimator \cite{kraskov2004estimating}. As each phase is defined on the circle $\theta \in [0, 2\pi)$, the raw phase values $\theta_i(t)$ were mapped onto $\mathbb{R}^2$ by $\mathbf{x}_i(t) = \bigl(\cos\theta_i(t),\;\sin\theta_i(t)\bigr)$. The last 80\% of each simulation was retained for analysis, matching the stable window used for PLV computation.

For each structurally connected edge $(i,j)$, mutual information $I(\mathbf{x}_i;\mathbf{x}_j)$ was estimated with the KSG variant~1 algorithm using the $L^\infty$ (Chebyshev) norm. For $T$ samples and $k$ nearest neighbors, the estimator is:

\begin{equation}
    \hat{I}(\mathbf{x}_i;\mathbf{x}_j) = \psi(k) + \psi(T) - \frac{1}{T}\sum_{t=1}^{T}\bigl[\psi(n_x(t)+1) + \psi(n_y(t)+1)\bigr] ~,
\end{equation}

\noindent where $\psi(\cdot)$ is the digamma function, and $n_x(t)$ and $n_y(t)$ are the number of points within the $L^\infty$ distance to the $k$-th neighbor in the joint space, projected onto the $\mathbf{x}_i$ and $\mathbf{x}_j$ marginals, respectively. The neighborhood parameter was set to $k = 5$, and a small jitter ($10^{-10}$) was added to break ties. Negative estimates were clipped to zero. Edge-wise MI computation was parallelized across five CPU cores.

\subsection{Edge-centric controllability}

Edge-level controllability was computed on the line graph (or edge-to-vertex dual) of the structural connectome (Fig. \ref{Fig1}B) \cite{faskowitz2020edge, sun2024edgecentric}. Given a weighted adjacency matrix $\mathbf{A} \in \mathbb{R}^{N \times N}$ with $M$ structurally connected edges (i.e., pairs with nonzero weight), a weighted incidence matrix $\mathbf{C} \in \mathbb{R}^{N \times M}$ was constructed such that $C_{i,a} = \sqrt{w_a}$ if node $i$ is an endpoint of edge $a$ (with weight $w_a$), and zero otherwise. The line graph adjacency was then obtained as:

\begin{equation}
    \mathbf{A}_E = \mathbf{C}^{\!\top}\mathbf{C} - \mathbf{W} ~,
\end{equation}

\noindent where $\mathbf{W} = \mathrm{diag}(\mathbf{w})$ removes self-loops. The resulting matrix was symmetrized and the negative entries were clipped to zero. To ensure stability of discrete-time controllability proxies, $\mathbf{A}_E$ was spectrally normalized to $\tilde{\mathbf{A}}_E = \mathbf{A}_E / (1 + \sigma_{\max}(\mathbf{A}_E))$, where $\sigma_{\max}$ denotes the largest singular value \cite{gu2015controllability, pasqualetti2014controllability}.

Edge average controllability (eAC) and edge modal controllability (eMC) were derived from the symmetric eigendecomposition $\tilde{\mathbf{A}}_E = \mathbf{Q}\boldsymbol{\Lambda}\mathbf{Q}^{\!\top}$, where $\boldsymbol{\Lambda} = \mathrm{diag}(\lambda_1,\dots,\lambda_M)$. For each edge $a$,

\begin{equation}
    \mathrm{eAC}_a = \sum_{l=1}^{M} \frac{Q_{a,l}^{2}}{1 - \lambda_l^{2}}, \qquad \mathrm{eMC}_a = \sum_{l=1}^{M} Q_{a,l}^{2}\,(1 - \lambda_l^{2}) ~,
\end{equation}

\noindent where $Q_{a,l}$ is the element $(a,l)$ of the eigenvector matrix. eAC quantifies how effectively perturbing the edge $a$ can drive the network toward easy-to-reach dynamical states, whereas eMC quantifies its capacity to reach difficult-to-reach states\cite{gu2015controllability}. The eigenvalues were bound to the interval $[-1,1]$ to prevent numerical divergence.

\subsection{QUIET edge ranking}

QUIET integrates the structural controllability and information features described above into a single, composite edge-level score. For each edge $a$ in the target sub-network, a control term and an information term are computed. The control term averages the percentile ranks of eAC and eMC:

\begin{equation}
    S_{\mathrm{ctrl}}(a) = \frac{\mathrm{rank}_{\%}(\mathrm{eAC}_a) + \mathrm{rank}_{\%}(\mathrm{eMC}_a)}{2} ~.
\end{equation}

\noindent The information term inverts the percentile-ranked mutual information, so that functionally underutilized edges, those with low MI and therefore low redundancy, receive higher scores:

\begin{equation}
    S_{\mathrm{info}}(a) = 1 - \mathrm{rank}_{\%}(\mathrm{MI}_a) ~.
\end{equation}

\noindent The final QUIET score is a weighted combination of the two terms:

\begin{equation}
    \mathrm{QUIET}(a) = w_{\mathrm{ctrl}}\,S_{\mathrm{ctrl}}(a) \;+\; w_{\mathrm{info}}\,S_{\mathrm{info}}(a) ~,
\end{equation}

\noindent where $w_{\mathrm{ctrl}}$ and $w_{\mathrm{info}}$ are non-negative weights normalized to sum to one. Critically, these weights are not fixed hyperparameters but are jointly optimized with the coupling parameters by the optimizer (see \ref{Optimization}), allowing the framework to autonomously discover the balance between structural influence and functional redundancy for each network. Edges are sorted by descending QUIET score, and the top-ranked edges form the candidate perturbation set. 

\subsection{Three-stage optimization pipeline}
\label{Optimization}

QUIET determines the optimal perturbation through a three-stage optimization pipeline, each stage formulated as a multi-objective or single-objective problem solved by a tree-structured Parzen estimator (TPE)\cite{bergstra2011algorithms}. At each optimization iteration, the phase oscillator system is simulated with the candidate edge set and synchronization is evaluated as described above. We define two performance metrics: the target synchronization $\bar{R}_{\mathrm{target}}$, computed as the mean PLV across all edges within the designated target sub-network, and the rest stability $\mathrm{MAE}_{\mathrm{rest}}$, computed as the mean absolute deviation between the current and baseline PLV matrices restricted to non-target nodes.

\subsubsection{Stage 1: Maximum synchronization}

The objective of Stage~1 is to establish the best achievable synchronization of the targeted region while minimizing off-target disturbances, without regard to energy cost. This stage does not yield the final perturbation; instead, it sets the synchronization target that Stages~2 and~3 must satisfy at minimal energy. A multi-objective TPE jointly optimizes the ranking weights ($w_{\mathrm{ctrl}}, w_{\mathrm{info}} \in [0, 1]$, normalized to sum to one), the coupling strength $\alpha \in [0.01, 1000]$, and the number of perturbed edges $n$ (integer in $[1, M]$, where $M$ is the number of structurally connected edges in the target sub-network). The two objectives are to maximize target synchronization and minimize off-target disruption:

\begin{equation}
    \min_{\,w_{\mathrm{ctrl}},\, w_{\mathrm{info}},\, \alpha,\, n}\;\bigl(-\bar{R}_{\mathrm{target}},\;\mathrm{MAE}_{\mathrm{rest}}\bigr) ~.
\end{equation}

\noindent The resulting Pareto front is evaluated, and the optimal operating point is selected as the iteration closest to the ideal corner ($\bar{R}_{\mathrm{target}} = 1$, $\mathrm{MAE}_{\mathrm{rest}} = 0$) in Euclidean distance. This ``sweet-spot'' optimization iteration defines the synchronization target for subsequent stages: $R^{*} = 0.95\,\bar{R}_{\mathrm{target}}^{\mathrm{sweet}}$ (a 5\% margin to ensure feasibility) and the rest tolerance $\delta^{*} = \max(0.005,\,\mathrm{MAE}_{\mathrm{rest}}^{\mathrm{sweet}})$.

\subsubsection{Stage 2: Energy minimization}

With the synchronization requirement fixed at $R^{*}$ from Stage~1, a second multi-objective TPE re-explores the ranking weights and coupling strength to minimize the required control energy. For each TPE iteration, i.e., for each candidate combination of ($w_{\mathrm{ctrl}}, w_{\mathrm{info}}, \alpha$), a binary search over the ranked edge list determines the minimum number of edges $n^{*}$ that satisfies $\bar{R}_{\mathrm{target}} \geq R^{*}$ and $\mathrm{MAE}_{\mathrm{rest}} \leq \delta^{*}$. The resulting pair ($n^{*}, \alpha$) is then returned to the TPE as the two objectives to minimize:

\begin{equation}
    \min_{\,w_{\mathrm{ctrl}},\, w_{\mathrm{info}},\, \alpha}\;\bigl(n^{*},\;\alpha\bigr) ~.
\end{equation}

\noindent The Pareto-optimal iteration that minimizes total control energy $E = n^{*} \times \alpha$ is selected. Here, we define the control energy ($E$) as the total magnitude of coupling-weight modulation aggregated across the perturbed edges, distinguishing it from the integrated input norm $\int_0^T \|u(t)\|^2\, dt$ of classical NCT.

\subsubsection{Stage 3: Coupling refinement}

Since the TPE in Stage~2 samples $\alpha$ stochastically across a wide range, it may not converge on the exact minimum coupling for the selected edge set. Stage~3 addresses this potential issue by holding the edge set from Stage~2 fixed and performing a deterministic binary search over $\alpha \in [0.001,\,\alpha_{\mathrm{Stage\,2}}]$ for 100 iterations to converge on the lowest coupling strength that still satisfies the synchronization and stability constraints. The final configuration $(n^{*}_{\mathrm{final}},\,\alpha_{\mathrm{final}})$ defines the QUIET-optimized perturbation, with total control energy $E = n^{*}_{\mathrm{final}} \times \alpha_{\mathrm{final}}$.

\subsubsection{Stages 1-3: Simulation details}

Each stage used 1000 TPE iterations with five parallel workers, and all simulations were run with identical initial phases, natural frequencies, and network topology to ensure comparability between optimization iterations.

\subsection{Statistical validation}

The significance of the QUIET-optimized edge set was assessed by comparison against a null distribution of random edge selections. For each experiment, the optimized number of edges $n^{*}$ and the coupling strength $\alpha^{*}$ were held fixed, and 10,000 random permutations of the candidate edge pool were drawn. Each permutation selected $n^{*}$ edges at random, applied the same coupling perturbation $\alpha^{*}$, and simulated the phase oscillator dynamics under identical initial conditions. Target synchronization was recorded for each permutation, yielding a null distribution of $\bar{R}_{\mathrm{null}}$ values. The synchronization $z$-score was computed as

\begin{equation}
    z_{\mathrm{sync}} = \frac{\bar{R}_{\mathrm{QUIET}} - \mu_{\mathrm{null}}}{\sigma_{\mathrm{null}}} ~,
\end{equation}

\noindent where $\bar{R}_{\mathrm{QUIET}}$ is the synchronization achieved by the QUIET-ranked edges and $\mu_{\mathrm{null}}$ and $\sigma_{\mathrm{null}}$ are the mean and standard deviation of the null distribution. The $p$-value was computed as $(s + 1)/(N_{\mathrm{perm}} + 1)$, where $s$ is the number of null permutations that matched or exceeded the QUIET performance on both synchronization and rest stability criteria simultaneously.

\section*{Citation diversity statement}

Recent work in several fields of science has identified a bias in citation practices such that papers from women and other minorities are under-cited relative to the number of such papers in the field \cite{mitchell2013gendered, maliniak2013gender, caplar2017quantitative, dion2018gendered, dworkin2020extent, bertolero2021racial, wang2021gendered, chatterjee2021gender, fulvio2021imbalance}. Here we sought to proactively consider choosing references that reflect the diversity of the field in thought, form of contribution, gender, race, ethnicity, and other factors. First, we obtained the predicted gender of the first and last author of each reference by querying public author records (lab pages, ORCID, and institutional profiles); ambiguous cases were verified manually \cite{dworkin2020extent, zhou2020gender}. By this measure (and excluding self-citations to the first author of our current paper), our references contain 5.4\% woman(first)/woman(last), 9.8\% man/woman, 16.3\% woman/man, and 54.3\% man/man. Second, we obtained the predicted racial/ethnic category of the first and last author of each reference by applying a probabilistic name classification model trained on Florida voter registration data, with race probabilities computed per-author and combined into per-paper bucket probabilities following the cleanBib protocol \cite{ambekar2009name, sood2018predicting, bertolero2021racial, zhou2020gender}. By this measure (and excluding self-citations), our references contain 16.3\% author of color(first)/author of color(last), 15.6\% white author/author of color, 23.5\% author of color/white author, and 44.6\% white/white. These methods are limited in that (a) names, pronouns, and public profiles may not, in every case, be indicative of gender or racial/ethnic identity; (b) the gender categorization cannot fully account for intersex, non-binary, or transgender people who use binary pronouns; and (c) the race classifier was trained on US data and may misclassify authors from underrepresented or mixed-race backgrounds, or those whose names face differential biases due to ambiguous racialization or ethnicization. We look forward to future work that could help us to better understand how to support equitable practices in science.

\pagebreak

\section*{Data availability}
The Human Connectome Project S1200 data analysed in this study are publicly available from the WU-Minn HCP Consortium through the ConnectomeDB repository (\url{https://db.humanconnectome.org}) under the HCP Open Access Data Use Terms. The Schaefer 400-region parcellation is publicly available \cite{schaefer2018local}. The University of Pennsylvania sample dataset analyzed as well as the scripts generated for the purposes of this study are available from the corresponding authors upon request. Source data are provided with this paper.

\section*{Code availability}
The Network Synchronization application is available as open-source software at (\url{https://soveshmohapatra.com/research/QUIET}) under the MIT license.

\bibliography{ref}

@article{gu2018energy,
  title     = {The energy landscape of neurophysiological activity implicit in brain network structure},
  author    = {Gu, Shi and Betzel, Richard F and Tang, Evelyn and Avena-Koenigsberger, Andrea and Pasqualetti, Fabio and Bassett, Danielle S},
  journal   = {Scientific Reports},
  volume    = {8},
  pages     = {2507},
  year      = {2018},
  doi       = {10.1038/s41598-018-20123-8},
  publisher = {Nature Publishing Group}
}

@article{baum2020development,
  title     = {Development of structure-function coupling in human brain networks during youth},
  author    = {Baum, Graham L and Cui, Zaixu and Roalf, David R and Ciric, Rastko and Betzel, Richard F and Larsen, Bart and Cieslak, Matthew and Cook, Philip A and Xia, Cedric H and Moore, Tyler M and Ruparel, Kosha and Oathes, Desmond J and Alexander-Bloch, Aaron F and Shinohara, Russell T and Raznahan, Armin and Gur, Raquel E and Gur, Ruben C and Bassett, Danielle S and Satterthwaite, Theodore D},
  journal   = {Proceedings of the National Academy of Sciences},
  volume    = {117},
  number    = {1},
  pages     = {771--778},
  year      = {2020},
  doi       = {10.1073/pnas.1912034117},
  publisher = {National Academy of Sciences}
}

@article{shanmugan2022sex,
  title     = {Sex differences in the functional topography of association networks in youth},
  author    = {Shanmugan, Sheila and Seidlitz, Jakob and Cui, Zaixu and Adebimpe, Azeez and Bassett, Danielle S and Bertolero, Maxwell A and Davatzikos, Christos and Fair, Damien A and Gur, Raquel E and Gur, Ruben C and Larsen, Bart and Li, Hongming and Pines, Adam and Raznahan, Armin and Roalf, David R and Shinohara, Russell T and Vogel, Jacob and Wolf, Daniel H and Fan, Yong and Alexander-Bloch, Aaron and Satterthwaite, Theodore D},
  journal   = {Proceedings of the National Academy of Sciences},
  volume    = {119},
  number    = {33},
  pages     = {e2110416119},
  year      = {2022},
  doi       = {10.1073/pnas.2110416119},
  publisher = {National Academy of Sciences}
}

@article{vazquez2019gradients,
  title     = {Gradients of structure-function tethering across neocortex},
  author    = {V{\'a}zquez-Rodr{\'i}guez, Bertha and Su{\'a}rez, Laura E and Markello, Ross D and Shafiei, Golia and Paquola, Casey and Hagmann, Patric and van den Heuvel, Martijn P and Bernhardt, Boris C and Spreng, R Nathan and Misic, Bratislav},
  journal   = {Proceedings of the National Academy of Sciences},
  volume    = {116},
  number    = {42},
  pages     = {21219--21227},
  year      = {2019},
  doi       = {10.1073/pnas.1903403116},
  publisher = {National Academy of Sciences}
}

@article{bassett2017network,
  title     = {Network neuroscience},
  author    = {Bassett, Danielle S and Sporns, Olaf},
  journal   = {Nature Neuroscience},
  volume    = {20},
  number    = {3},
  pages     = {353--364},
  year      = {2017},
  doi       = {10.1038/nn.4502},
  publisher = {Nature Publishing Group}
}

@article{bullmore2009complex,
  title     = {Complex brain networks: graph theoretical analysis of structural and functional systems},
  author    = {Bullmore, Ed and Sporns, Olaf},
  journal   = {Nature Reviews Neuroscience},
  volume    = {10},
  number    = {3},
  pages     = {186--198},
  year      = {2009},
  doi       = {10.1038/nrn2575},
  publisher = {Nature Publishing Group}
}

@article{honey2009predicting,
  title     = {Predicting human resting-state functional connectivity from structural connectivity},
  author    = {Honey, Christopher J and Sporns, Olaf and Cammoun, Leila and Gigandet, Xavier and Thiran, Jean-Philippe and Meuli, Reto and Hagmann, Patric},
  journal   = {Proceedings of the National Academy of Sciences},
  volume    = {106},
  number    = {6},
  pages     = {2035--2040},
  year      = {2009},
  doi       = {10.1073/pnas.0811168106},
  publisher = {National Academy of Sciences}
}

@article{deco2011emerging,
  title     = {Emerging concepts for the dynamical organization of resting-state activity in the brain},
  author    = {Deco, Gustavo and Jirsa, Viktor K and McIntosh, Anthony R},
  journal   = {Nature Reviews Neuroscience},
  volume    = {12},
  number    = {1},
  pages     = {43--56},
  year      = {2011},
  doi       = {10.1038/nrn2961},
  publisher = {Nature Publishing Group}
}

@article{tang2017developmental,
  title     = {Developmental increases in white matter network controllability support a growing diversity of brain dynamics},
  author    = {Tang, Evelyn and Giusti, Chad and Baum, Graham L and Gu, Shi and Pollock, Eli and Kahn, Ari E and Roalf, David R and Moore, Tyler M and Ruparel, Kosha and Gur, Ruben C and Gur, Raquel E and Satterthwaite, Theodore D and Bassett, Danielle S},
  journal   = {Nature Communications},
  volume    = {8},
  number    = {1},
  pages     = {1252},
  year      = {2017},
  doi       = {10.1038/s41467-017-01254-4},
  publisher = {Nature Publishing Group}
}

@article{pasqualetti2014controllability,
  title     = {Controllability metrics, limitations and algorithms for complex networks},
  author    = {Pasqualetti, Fabio and Zampieri, Sandro and Bullo, Francesco},
  journal   = {IEEE Transactions on Control of Network Systems},
  volume    = {1},
  number    = {1},
  pages     = {40--52},
  year      = {2014},
  doi       = {10.1109/TCNS.2014.2310254},
  publisher = {IEEE}
}

@article{gu2015controllability,
  title     = {Controllability of structural brain networks},
  author    = {Gu, Shi and Pasqualetti, Fabio and Cieslak, Matthew and Telesford, Qawi K and Yu, Alfred B and Kahn, Ari E and Medaglia, John D and Vettel, Jean M and Miller, Michael B and Grafton, Scott T and Bassett, Danielle S},
  journal   = {Nature Communications},
  volume    = {6},
  number    = {1},
  pages     = {8414},
  year      = {2015},
  doi       = {10.1038/ncomms9414},
  publisher = {Nature Publishing Group}
}

@article{tang2018control,
  title     = {Colloquium: Control of dynamics in brain networks},
  author    = {Tang, Evelyn and Bassett, Danielle S},
  journal   = {Reviews of Modern Physics},
  volume    = {90},
  number    = {3},
  pages     = {031003},
  year      = {2018},
  doi       = {10.1103/RevModPhys.90.031003},
  publisher = {American Physical Society}
}

@article{avena2018communication,
  title     = {Communication dynamics in complex brain networks},
  author    = {Avena-Koenigsberger, Andrea and Mišić, Bratislav and Sporns, Olaf},
  journal   = {Nature Reviews Neuroscience},
  volume    = {19},
  number    = {1},
  pages     = {17--33},
  year      = {2018},
  doi       = {10.1038/nrn.2017.149},
  publisher = {Nature Publishing Group}
}

@article{vandenheuvel2011rich,
  title     = {Rich-club organization of the human connectome},
  author    = {van den Heuvel, Martijn P and Sporns, Olaf},
  journal   = {Journal of Neuroscience},
  volume    = {31},
  number    = {44},
  pages     = {15775--15786},
  year      = {2011},
  doi       = {10.1523/JNEUROSCI.3539-11.2011},
  publisher = {Society for Neuroscience}
}

@article{bullmore2012economy,
  title     = {The economy of brain network organization},
  author    = {Bullmore, Ed and Sporns, Olaf},
  journal   = {Nature Reviews Neuroscience},
  volume    = {13},
  number    = {5},
  pages     = {336--349},
  year      = {2012},
  doi       = {10.1038/nrn3214},
  publisher = {Nature Publishing Group}
}

@article{karrer2020practical,
  title     = {A practical guide to methodological considerations in the controllability of structural brain networks},
  author    = {Karrer, Teresa M and Kim, Jason Z and Stiso, Jennifer and Kahn, Ari E and Pasqualetti, Fabio and Habel, Ute and Bassett, Danielle S},
  journal   = {Journal of Neural Engineering},
  volume    = {17},
  number    = {2},
  pages     = {026031},
  year      = {2020},
  doi       = {10.1088/1741-2552/ab6e8b},
  publisher = {IOP Publishing}
}

@article{faskowitz2020edge,
  title     = {Edge-centric functional network representations of human cerebral cortex reveal overlapping system-level architecture},
  author    = {Faskowitz, Joshua and Esfahlani, Farnaz Zamani and Jo, Youngheun and Sporns, Olaf and Betzel, Richard F},
  journal   = {Nature Neuroscience},
  volume    = {23},
  number    = {12},
  pages     = {1644--1654},
  year      = {2020},
  doi       = {10.1038/s41593-020-00719-y},
  publisher = {Nature Publishing Group}
}

@article{braun2021brain,
  title     = {Brain network dynamics during working memory are modulated by dopamine and diminished in schizophrenia},
  author    = {Braun, Urs and Harneit, Anais and Pergola, Giulio and Menara, Tommaso and Schaefer, Axel and Betzel, Richard F and Zang, Zhenxiang and Schweiger, Janina I and Zhang, Xiaolong and Schwarz, Karsten and Chen, Junfang and Blasi, Giuseppe and Bertolino, Alessandro and Durstewitz, Daniel and Pasqualetti, Fabio and Schwarz, Emanuel and Meyer-Lindenberg, Andreas and Bassett, Danielle S and Tost, Heike},
  journal   = {Nature Communications},
  volume    = {12},
  number    = {1},
  pages     = {3478},
  year      = {2021},
  doi       = {10.1038/s41467-021-23694-9},
  publisher = {Nature Publishing Group}
}

@article{stiso2019white,
  title     = {White matter network architecture guides direct electrical stimulation through optimal state transitions},
  author    = {Stiso, Jennifer and Khambhati, Ankit N and Menara, Tommaso and Kahn, Ari E and Stein, Joel M and Das, Sandhitsu R and Gorniak, Richard and Tracy, Joseph and Litt, Brian and Davis, Kathryn A and Pasqualetti, Fabio and Lucas, Timothy H and Bassett, Danielle S},
  journal   = {Cell Reports},
  volume    = {28},
  number    = {10},
  pages     = {2554--2566},
  year      = {2019},
  doi       = {10.1016/j.celrep.2019.08.008},
  publisher = {Elsevier}
}

@article{medaglia2015cognitive,
  title     = {Cognitive network neuroscience},
  author    = {Medaglia, John D and Lynall, Mary-Ellen and Bassett, Danielle S},
  journal   = {Journal of Cognitive Neuroscience},
  volume    = {27},
  number    = {8},
  pages     = {1471--1491},
  year      = {2015},
  doi       = {10.1162/jocn\_a\_00810},
  publisher = {MIT Press}
}

@article{timme2018tutorial,
  title     = {A tutorial for information theory in neuroscience},
  author    = {Timme, Nicholas M and Lapish, Christopher},
  journal   = {eNeuro},
  volume    = {5},
  number    = {3},
  pages     = {ENEURO.0052-18.2018},
  year      = {2018},
  doi       = {10.1523/ENEURO.0052-18.2018},
  publisher = {Society for Neuroscience}
}

@article{kraskov2004estimating,
  title     = {Estimating mutual information},
  author    = {Kraskov, Alexander and St{\"o}gbauer, Harald and Grassberger, Peter},
  journal   = {Physical Review E},
  volume    = {69},
  number    = {6},
  pages     = {066138},
  year      = {2004},
  doi       = {10.1103/PhysRevE.69.066138},
  publisher = {American Physical Society}
}

@inproceedings{bergstra2011algorithms,
  title     = {Algorithms for hyper-parameter optimization},
  author    = {Bergstra, James and Bardenet, R{\'e}mi and Bengio, Yoshua and K{\'e}gl, Bal{\'a}zs},
  booktitle = {Advances in Neural Information Processing Systems},
  volume    = {24},
  year      = {2011},
  publisher = {Curran Associates}
}

@article{acebron2005kuramoto,
  title     = {The {Kuramoto} model: A simple paradigm for synchronization phenomena},
  author    = {Acebrón, Juan A and Bonilla, L L and P{\'e}rez Vicente, Conrad J and Ritort, F{\'e}lix and Spigler, Renato},
  journal   = {Reviews of Modern Physics},
  volume    = {77},
  number    = {1},
  pages     = {137--185},
  year      = {2005},
  doi       = {10.1103/RevModPhys.77.137},
  publisher = {American Physical Society}
}

@book{kuramoto1984chemical,
  title     = {Chemical Oscillations, Waves, and Turbulence},
  author    = {Kuramoto, Yoshiki},
  year      = {1984},
  publisher = {Springer-Verlag},
  address   = {Berlin},
  doi       = {10.1007/978-3-642-69689-3}
}

@article{lachaux1999measuring,
  title     = {Measuring phase synchrony in brain signals},
  author    = {Lachaux, Jean-Philippe and Rodriguez, Eugenio and Martinerie, Jacques and Varela, Francisco J},
  journal   = {Human Brain Mapping},
  volume    = {8},
  number    = {4},
  pages     = {194--208},
  year      = {1999},
  doi       = {10.1002/(SICI)1097-0193(1999)8:4<194::AID-HBM4>3.0.CO;2-C},
  publisher = {Wiley}
}

@article{liu2011controllability,
  title     = {Controllability of complex networks},
  author    = {Liu, Yang-Yu and Slotine, Jean-Jacques and Barabási, Albert-László},
  journal   = {Nature},
  volume    = {473},
  number    = {7346},
  pages     = {167--173},
  year      = {2011},
  doi       = {10.1038/nature10011},
  publisher = {Nature Publishing Group}
}

@article{yan2012controlling,
  title     = {Controlling complex networks: How much energy is needed?},
  author    = {Yan, Gang and Ren, Jie and Lai, Ying-Cheng and Lai, Choy-Heng and Li, Baowen},
  journal   = {Physical Review Letters},
  volume    = {108},
  number    = {21},
  pages     = {218703},
  year      = {2012},
  doi       = {10.1103/PhysRevLett.108.218703},
  publisher = {American Physical Society}
}

@article{vanessen2013wuminn,
  title     = {The {WU-Minn} Human Connectome Project: An overview},
  author    = {Van Essen, David C and Smith, Stephen M and Barch, Deanna M and Behrens, Timothy EJ and Yacoub, Essa and Ugurbil, Kamil and {WU-Minn HCP Consortium}},
  journal   = {NeuroImage},
  volume    = {80},
  pages     = {62--79},
  year      = {2013},
  doi       = {10.1016/j.neuroimage.2013.05.041},
  publisher = {Elsevier}
}

@article{schaefer2018local,
  title     = {Local-global parcellation of the human cerebral cortex from intrinsic functional connectivity {MRI}},
  author    = {Schaefer, Alexander and Kong, Ru and Gordon, Evan M and Laumann, Timothy O and Zuo, Xi-Nian and Holmes, Avram J and Eickhoff, Simon B and Yeo, B T Thomas},
  journal   = {Cerebral Cortex},
  volume    = {28},
  number    = {9},
  pages     = {3095--3114},
  year      = {2018},
  doi       = {10.1093/cercor/bhx179},
  publisher = {Oxford University Press}
}

@article{seeley2007dissociable,
  title     = {Dissociable intrinsic connectivity networks for salience processing and executive control},
  author    = {Seeley, William W and Menon, Vinod and Schatzberg, Alan F and Keller, Jennifer and Glover, Gary H and Kenna, Heather and Reiss, Allan L and Greicius, Michael D},
  journal   = {Journal of Neuroscience},
  volume    = {27},
  number    = {9},
  pages     = {2349--2356},
  year      = {2007},
  doi       = {10.1523/JNEUROSCI.5587-06.2007},
  publisher = {Society for Neuroscience}
}

@article{menon2010saliency,
  title     = {Saliency, switching, attention and control: A network model of insula function},
  author    = {Menon, Vinod and Uddin, Lucina Q},
  journal   = {Brain Structure and Function},
  volume    = {214},
  number    = {5--6},
  pages     = {655--667},
  year      = {2010},
  doi       = {10.1007/s00429-010-0262-0},
  publisher = {Springer}
}

@article{uddin2015salience,
  title     = {Salience processing and insular cortical function and dysfunction},
  author    = {Uddin, Lucina Q},
  journal   = {Nature Reviews Neuroscience},
  volume    = {16},
  number    = {1},
  pages     = {55--61},
  year      = {2015},
  doi       = {10.1038/nrn3857},
  publisher = {Nature Publishing Group}
}

@article{raichle2015brain,
  title     = {The brain's default mode network},
  author    = {Raichle, Marcus E},
  journal   = {Annual Review of Neuroscience},
  volume    = {38},
  pages     = {433--447},
  year      = {2015},
  doi       = {10.1146/annurev-neuro-071013-014030},
  publisher = {Annual Reviews}
}

@article{whitfieldgabrieli2012default,
  title     = {Default mode network activity and connectivity in psychopathology},
  author    = {Whitfield-Gabrieli, Susan and Ford, Judith M},
  journal   = {Annual Review of Clinical Psychology},
  volume    = {8},
  pages     = {49--76},
  year      = {2012},
  doi       = {10.1146/annurev-clinpsy-032511-143049},
  publisher = {Annual Reviews}
}

@article{uddin2011dynamic,
  title     = {Dynamic reconfiguration of structural and functional connectivity across core neurocognitive brain networks with development},
  author    = {Uddin, Lucina Q and Supekar, Kaustubh and Menon, Vinod},
  journal   = {Journal of Neuroscience},
  volume    = {31},
  number    = {50},
  pages     = {18578--18589},
  year      = {2011},
  doi       = {10.1523/JNEUROSCI.4465-11.2011},
  publisher = {Society for Neuroscience}
}

@article{anticevic2012role,
  title     = {The role of default network deactivation in cognition and disease},
  author    = {Anticevic, Alan and Cole, Michael W and Murray, John D and Corlett, Philip R and Wang, Xiao-Jing and Krystal, John H},
  journal   = {Trends in Cognitive Sciences},
  volume    = {16},
  number    = {12},
  pages     = {584--592},
  year      = {2012},
  doi       = {10.1016/j.tics.2012.10.008},
  publisher = {Elsevier}
}

@article{fotiadis2025changes,
  title     = {Changes in brain connectivity and neurovascular dynamics during dexmedetomidine-induced loss of consciousness},
  author    = {Fotiadis, Panagiotis and McKinstry-Wu, Andrew R and Weinstein, Sarah M and Cook, Philip A and Elliott, Mark and Cieslak, Matthew and Duda, Jeffrey T and Satterthwaite, Theodore D and Shinohara, Russell T and Proekt, Alexander and Kelz, Max B and Detre, John A and Bassett, Danielle S},
  journal   = {Communications Biology},
  volume    = {8},
  number    = {1},
  pages     = {1--15},
  year      = {2025},
  doi       = {10.1038/s42003-025-08577-9},
  publisher = {Nature Publishing Group}
}

@article{boveroux2010breakdown,
  title     = {Breakdown of within- and between-network resting state functional magnetic resonance imaging connectivity during propofol-induced loss of consciousness},
  author    = {Boveroux, Pierre and Vanhaudenhuyse, Audrey and Bruno, Marie-Aur{\'e}lie and Noirhomme, Quentin and Lauwick, S{\'e}bastien and Luxen, Andr{\'e} and Degueldre, Christian and Plenevaux, Alain and Schnakers, Caroline and Phillips, Christophe and Brichant, Jean-Fran{\c{c}}ois and Bonhomme, Vincent and Maquet, Pierre and Greicius, Michael D and Laureys, Steven and Boly, M{\'e}lanie},
  journal   = {Anesthesiology},
  volume    = {113},
  number    = {5},
  pages     = {1038--1053},
  year      = {2010},
  doi       = {10.1097/ALN.0b013e3181f697f5},
  publisher = {Wolters Kluwer}
}

@article{hashmi2017dexmedetomidine,
  title     = {Dexmedetomidine disrupts the local and global efficiencies of large-scale brain networks},
  author    = {Hashmi, Javeria A and Loggia, Marco L and Khan, Sheraz and Gao, Lei and Kim, Jieun and Napadow, Vitaly and Brown, Emery N and Akeju, Oluwaseun},
  journal   = {Anesthesiology},
  volume    = {126},
  number    = {3},
  pages     = {419--430},
  year      = {2017},
  doi       = {10.1097/ALN.0000000000001509},
  publisher = {Wolters Kluwer}
}

@article{demertzi2019human,
  title     = {Human consciousness is supported by dynamic complex patterns of brain signal coordination},
  author    = {Demertzi, Athena and Tagliazucchi, Enzo and Dehaene, Stanislas and Deco, Gustavo and Barttfeld, Pablo and Raimondo, Federico and Martial, Charlotte and Fern{\'a}ndez-Espejo, Davinia and Rohaut, Benjamin and Voss, Henning U and Schiff, Nicholas D and Owen, Adrian M and Laureys, Steven and Naccache, Lionel and Sitt, Jacobo D},
  journal   = {Science Advances},
  volume    = {5},
  number    = {2},
  pages     = {eaat7603},
  year      = {2019},
  doi       = {10.1126/sciadv.aat7603},
  publisher = {American Association for the Advancement of Science}
}

@article{luppi2019consciousness,
  title     = {Consciousness-specific dynamic interactions of brain integration and functional diversity},
  author    = {Luppi, Andrea I and Craig, Michael M and Pappas, Ioannis and Finoia, Paola and Williams, Guy B and Allanson, Judith and Pickard, John D and Owen, Adrian M and Naci, Lorina and Menon, David K and Stamatakis, Emmanuel A},
  journal   = {Nature Communications},
  volume    = {10},
  number    = {1},
  pages     = {4616},
  year      = {2019},
  doi       = {10.1038/s41467-019-12658-9},
  publisher = {Nature Publishing Group}
}

@article{buchanan2009patient,
  title     = {Patient sex and its influence on general anaesthesia},
  author    = {Buchanan, Fiona F and Myles, Paul S and Cicuttini, Flavia},
  journal   = {Anaesthesia and Intensive Care},
  volume    = {37},
  number    = {2},
  pages     = {207--218},
  year      = {2009},
  doi       = {10.1177/0310057X0903700201},
  publisher = {SAGE Publications}
}

@article{maier2017challenge,
  title     = {The challenge of mapping the human connectome based on diffusion tractography},
  author    = {Maier-Hein, Klaus H and Neher, Peter F and Houde, Jean-Christophe and C{\^o}t{\'e}, Marc-Alexandre and Garyfallidis, Eleftherios and Zhong, Jidan and Chamberland, Maxime and Yeh, Fang-Cheng and Lin, Ying-Chia and Ji, Qing and others},
  journal   = {Nature Communications},
  volume    = {8},
  number    = {1},
  pages     = {1349},
  year      = {2017},
  doi       = {10.1038/s41467-017-01285-x},
  publisher = {Nature Publishing Group}
}

@article{fotiadis2023myelination,
  title     = {Myelination and excitation-inhibition balance synergistically shape structure-function coupling across the human cortex},
  author    = {Fotiadis, Panagiotis and Cieslak, Matthew and He, Xiaosong and Caciagli, Lorenzo and Ouellet, Mathieu and Satterthwaite, Theodore D and Shinohara, Russell T and Bassett, Danielle S},
  journal   = {Nature Communications},
  volume    = {14},
  number    = {1},
  pages     = {6115},
  year      = {2023},
  doi       = {10.1038/s41467-023-41686-9},
  publisher = {Nature Publishing Group}
}

@InProceedings{MohSov_TReND_MICCAI2025,
  author = {Mohapatra, Sovesh AND Ouyang, Minhui AND Tan, Shufang AND Guo, Jianlin AND Sun, Lianglong AND He, Yong AND Huang, Hao},
  title = {TReND: Transformer derived features and Regularized NMF for neonatal functional network Delineation},
  booktitle = {Proceedings of Medical Image Computing and Computer Assisted Intervention -- MICCAI 2025},
  year = {2025},
  publisher = {Springer Nature Switzerland},
  volume = {LNCS 15971},
  month = {September},
  page = {660 -- 669},
}

@article{glasser2013minimal,
  title={The minimal preprocessing pipelines for the {Human Connectome Project}},
  author={Glasser, Matthew F and Sotiropoulos, Stamatios N and Wilson, J Anthony and Coalson, Timothy S and Fischl, Bruce and Andersson, Jesper L and Xu, Junqian and Jbabdi, Saad and Webster, Matthew and Polimeni, Jonathan R and others},
  journal={NeuroImage},
  volume={80},
  pages={105--124},
  year={2013},
  publisher={Elsevier},
  doi={10.1016/j.neuroimage.2013.04.127}
}

@article{tournier2019mrtrix3,
  title={{MRtrix3}: A fast, flexible and open software framework for medical image processing and visualisation},
  author={Tournier, J-Donald and Smith, Robert and Raffelt, David and Tabbara, Rami and Dhollander, Thijs and Pietsch, Maximilian and Christiaens, Daan and Jeurissen, Ben and Yeh, Chun-Hung and Connelly, Alan},
  journal={NeuroImage},
  volume={202},
  pages={116137},
  year={2019},
  publisher={Elsevier},
  doi={10.1016/j.neuroimage.2019.116137}
}

@article{smith2012anatomically,
  title={Anatomically-constrained tractography: improved diffusion {MRI} streamlines tractography through effective use of anatomical information},
  author={Smith, Robert E and Tournier, Jacques-Donald and Calamante, Fernando and Connelly, Alan},
  journal={NeuroImage},
  volume={62},
  number={3},
  pages={1924--1938},
  year={2012},
  publisher={Elsevier},
  doi={10.1016/j.neuroimage.2012.06.005}
}

@article{smith2015sift2,
  title={{SIFT2}: Enabling dense quantitative assessment of brain white matter connectivity using streamlines tractography},
  author={Smith, Robert E and Tournier, Jacques-Donald and Calamante, Fernando and Connelly, Alan},
  journal={NeuroImage},
  volume={119},
  pages={338--351},
  year={2015},
  publisher={Elsevier},
  doi={10.1016/j.neuroimage.2015.06.092}
}

@article{glasser2016multi,
  title={A multi-modal parcellation of human cerebral cortex},
  author={Glasser, Matthew F and Coalson, Timothy S and Robinson, Emma C and Hacker, Carl D and Harwell, John and Yacoub, Essa and Ugurbil, Kamil and Andersson, Jesper and Beckmann, Christian F and Jenkinson, Mark and others},
  journal={Nature},
  volume={536},
  number={7615},
  pages={171--178},
  year={2016},
  publisher={Nature Publishing Group},
  doi={10.1038/nature18933}
}

@article{salimi2014automatic,
  title={Automatic denoising of functional {MRI} data: combining independent component analysis and hierarchical fusion of classifiers},
  author={Salimi-Khorshidi, Gholamreza and Douaud, Gwena{\"e}lle and Beckmann, Christian F and Glasser, Matthew F and Griffanti, Ludovica and Smith, Stephen M},
  journal={NeuroImage},
  volume={90},
  pages={449--468},
  year={2014},
  publisher={Elsevier},
  doi={10.1016/j.neuroimage.2013.11.046}
}

@article{yeo2011organization,
  title={The organization of the human cerebral cortex estimated by intrinsic functional connectivity},
  author={Yeo, BT Thomas and Krienen, Fenna M and Sepulcre, Jorge and Sabuncu, Mert R and Lashkari, Danial and Hollinshead, Marisa and Roffman, Joshua L and Smoller, Jordan W and Z{\"o}llei, Lilla and Polimeni, Jonathan R and Fischl, Bruce and Liu, Hesheng and Buckner, Randy L},
  journal={Journal of Neurophysiology},
  volume={106},
  number={3},
  pages={1125--1165},
  year={2011},
  publisher={American Physiological Society},
  doi={10.1152/jn.00338.2011}
}

@article{sun2024edgecentric,
  title={Edge-centric network control on the human brain structural network},
  author={Sun, Huili and Deng, Haodong and Yan, Gang},
  journal={Imaging Neuroscience},
  volume={2},
  pages={1--17},
  year={2024},
  publisher={MIT Press},
  doi={10.1162/imag_a_00191}
}

@article{honey2007network,
  title={Network structure of cerebral cortex shapes functional connectivity on multiple time scales},
  author={Honey, Christopher J and K{\"o}tter, Rolf and Breakspear, Michael and Sporns, Olaf},
  journal={Proceedings of the National Academy of Sciences},
  volume={104},
  number={24},
  pages={10240--10245},
  year={2007},
  publisher={National Academy of Sciences},
  doi={10.1073/pnas.0701519104}
}

@article{cabral2011role,
  title={Role of local network oscillations in resting-state functional connectivity},
  author={Cabral, Joana and Hugues, Etienne and Sporns, Olaf and Deco, Gustavo},
  journal={NeuroImage},
  volume={57},
  number={1},
  pages={130--139},
  year={2011},
  publisher={Elsevier},
  doi={10.1016/j.neuroimage.2011.04.010}
}

@article{cornelius2013realistic,
  title     = {Realistic control of network dynamics},
  author    = {Cornelius, Sean P and Kath, William L and Motter, Adilson E},
  journal   = {Nature Communications},
  volume    = {4},
  number    = {1},
  pages     = {1942},
  year      = {2013},
  publisher = {Nature Publishing Group},
  doi       = {10.1038/ncomms2939}
}

@article{yan2017network,
  title     = {Network control principles predict neuron function in the {Caenorhabditis} elegans connectome},
  author    = {Yan, Gang and V{\'e}rtes, Petra E and Towlson, Emma K and Chew, Yee Lian and Walker, Denise S and Schafer, William R and Barab{\'a}si, Albert-L{\'a}szl{\'o}},
  journal   = {Nature},
  volume    = {550},
  number    = {7677},
  pages     = {519--523},
  year      = {2017},
  publisher = {Nature Publishing Group},
  doi       = {10.1038/nature24056}
}

@article{yan2011sex,
  title     = {Sex- and Brain Size--Related Small-World Structural Cortical Network in Young Adults: A {DTI} Tractography Study},
  author    = {Yan, Chaogan and Gong, Gaolang and Wang, Jinhui and Wang, Danyang and Liu, Dongqiang and Zhu, Chaozhe and Chen, Zhang J and Evans, Alan and Zang, Yufeng and He, Yong},
  journal   = {Cerebral Cortex},
  volume    = {21},
  number    = {2},
  pages     = {449--458},
  year      = {2011},
  publisher = {Oxford University Press},
  doi       = {10.1093/cercor/bhq111}
}

@article{singleton2022receptor,
  title     = {Receptor-informed network control theory links {LSD} and psilocybin to a flattening of the brain's control energy landscape},
  author    = {Singleton, S Parker and Luppi, Andrea I and Carhart-Harris, Robin L and Cruzat, Josephine and Roseman, Leor and Nutt, David J and Deco, Gustavo and Kringelbach, Morten L and Stamatakis, Emmanuel A and Kuceyeski, Amy},
  journal   = {Nature Communications},
  volume    = {13},
  number    = {1},
  pages     = {5812},
  year      = {2022},
  publisher = {Nature Publishing Group},
  doi       = {10.1038/s41467-022-33578-1}
}

@article{klickstein2017energy,
  title     = {Energy scaling of targeted optimal control of complex networks},
  author    = {Klickstein, Isaac and Shirin, Afroza and Sorrentino, Francesco},
  journal   = {Nature Communications},
  volume    = {8},
  number    = {1},
  pages     = {15145},
  year      = {2017},
  publisher = {Nature Publishing Group},
  doi       = {10.1038/ncomms15145}
}

@article{benmessaoud2025lowdim,
  title     = {Low-dimensional controllability of brain networks},
  author    = {Ben Messaoud, Remy and Le Du, Vincent and Bousfiha, Camile and Corsi, Marie-Constance and Gonzalez-Astudillo, Juliana and Kaufmann, Brigitte Charlotte and Venot, Tristan and Couvy-Duchesne, Baptiste and Migliaccio, Lara and Rosso, Charlotte and Bartolomeo, Paolo and Chavez, Mario and De Vico Fallani, Fabrizio},
  journal   = {PLOS Computational Biology},
  volume    = {21},
  number    = {1},
  pages     = {e1012691},
  year      = {2025},
  publisher = {Public Library of Science},
  doi       = {10.1371/journal.pcbi.1012691}
}

@article{beynel2020structural,
  title     = {Structural Controllability Predicts Functional Patterns and Brain Stimulation Benefits Associated with Working Memory},
  author    = {Beynel, Lysianne and Deng, Lifu and Crowell, Courtney A and Dannhauer, Moritz and Palmer, Hannah and Hilbig, Susan and Peterchev, Angel V and Luber, Bruce and Lisanby, Sarah H and Cabeza, Roberto and Appelbaum, Lawrence G and Davis, Simon W},
  journal   = {Journal of Neuroscience},
  volume    = {40},
  number    = {35},
  pages     = {6770--6778},
  year      = {2020},
  publisher = {Society for Neuroscience},
  doi       = {10.1523/JNEUROSCI.0531-20.2020}
}

@article{neymotin2020hnn,
  title     = {Human Neocortical Neurosolver: A new software tool for interpreting the cellular and network origin of human {MEG/EEG} data},
  author    = {Neymotin, Samuel A and Daniels, Dylan S and Caldwell, Blake and McDougal, Robert A and Carnevale, Nicholas T and Jas, Mainak and Moore, Christopher I and Hines, Michael L and H{\"a}m{\"a}l{\"a}inen, Matti and Jones, Stephanie R},
  journal   = {eLife},
  volume    = {9},
  pages     = {e51214},
  year      = {2020},
  publisher = {eLife Sciences Publications, Ltd},
  doi       = {10.7554/eLife.51214}
}

@article{tolley2024sbi,
  title     = {Methods and considerations for estimating parameters in biophysically detailed neural models with simulation based inference},
  author    = {Tolley, Nicholas and Rodrigues, Pedro L C and Gramfort, Alexandre and Jones, Stephanie R},
  journal   = {PLOS Computational Biology},
  volume    = {20},
  number    = {2},
  pages     = {e1011108},
  year      = {2024},
  publisher = {Public Library of Science},
  doi       = {10.1371/journal.pcbi.1011108}
}

@article{palmer2021maximally,
  title     = {Maximally efficient prediction in the early fly visual system may support evasive flight maneuvers},
  author    = {Wang, Siwei and Segev, Idan and Borst, Alexander and Palmer, Stephanie},
  journal   = {PLOS Computational Biology},
  volume    = {17},
  number    = {5},
  pages     = {e1008965},
  year      = {2021},
  publisher = {Public Library of Science},
  doi       = {10.1371/journal.pcbi.1008965}
}

@article{nozari2024macroscopic,
  title     = {Macroscopic resting-state brain dynamics are best described by linear models},
  author    = {Nozari, Erfan and Bertolero, Maxwell A and Stiso, Jennifer and Caciagli, Lorenzo and Cornblath, Eli J and He, Xiaosong and Mahadevan, Arun S and Pappas, George J and Bassett, Dani S},
  journal   = {Nature Biomedical Engineering},
  volume    = {8},
  number    = {1},
  pages     = {68--84},
  year      = {2024},
  publisher = {Nature Publishing Group},
  doi       = {10.1038/s41551-023-01117-y}
}

@article{cole2013multitask,
  title     = {Multi-task connectivity reveals flexible hubs for adaptive task control},
  author    = {Cole, Michael W and Reynolds, Jeremy R and Power, Jonathan D and Repovs, Grega and Anticevic, Alan and Braver, Todd S},
  journal   = {Nature Neuroscience},
  volume    = {16},
  number    = {9},
  pages     = {1348--1355},
  year      = {2013},
  publisher = {Nature Publishing Group},
  doi       = {10.1038/nn.3470}
}

@article{vidaurre2017brain,
  title     = {Brain network dynamics are hierarchically organized in time},
  author    = {Vidaurre, Diego and Smith, Stephen M and Woolrich, Mark W},
  journal   = {Proceedings of the National Academy of Sciences},
  volume    = {114},
  number    = {48},
  pages     = {12827--12832},
  year      = {2017},
  publisher = {National Academy of Sciences},
  doi       = {10.1073/pnas.1705120114}
}

@article{fox2012efficacy,
  title     = {Efficacy of transcranial magnetic stimulation targets for depression is related to intrinsic functional connectivity with the subgenual cingulate},
  author    = {Fox, Michael D and Buckner, Randy L and White, Matthew P and Greicius, Michael D and Pascual-Leone, Alvaro},
  journal   = {Biological Psychiatry},
  volume    = {72},
  number    = {7},
  pages     = {595--603},
  year      = {2012},
  publisher = {Elsevier},
  doi       = {10.1016/j.biopsych.2012.04.028}
}

@article{ozdemir2020individualized,
  title     = {Individualized perturbation of the human connectome reveals reproducible biomarkers of network dynamics relevant to cognition},
  author    = {Ozdemir, Recep A and Tadayon, Ehsan and Boucher, Pierre and Momi, Davide and Karakhanyan, Kelly A and Fox, Michael D and Halko, Mark A and Pascual-Leone, Alvaro and Shafi, Mouhsin M and Santarnecchi, Emiliano},
  journal   = {Proceedings of the National Academy of Sciences},
  volume    = {117},
  number    = {14},
  pages     = {8115--8125},
  year      = {2020},
  publisher = {National Academy of Sciences},
  doi       = {10.1073/pnas.1911240117}
}

@article{sun2023neonatal,
  title     = {Network controllability of structural connectomes in the neonatal brain},
  author    = {Sun, Huili and Jiang, Rongtao and Dai, Wei and Dufford, Alexander J and Noble, Stephanie and Spann, Marisa N and Gu, Shi and Scheinost, Dustin},
  journal   = {Nature Communications},
  volume    = {14},
  number    = {1},
  pages     = {5820},
  year      = {2023},
  publisher = {Nature Publishing Group},
  doi       = {10.1038/s41467-023-41499-w}
}

@article{bassignana2022aging,
  title     = {The impact of aging on human brain network target controllability},
  author    = {Bassignana, Giulia and Lacidogna, Giordano and Bartolomeo, Paolo and Colliot, Olivier and De Vico Fallani, Fabrizio},
  journal   = {Brain Structure and Function},
  volume    = {227},
  number    = {9},
  pages     = {3001--3015},
  year      = {2022},
  publisher = {Springer},
  doi       = {10.1007/s00429-022-02584-w}
}

@article{hahn2023genetic,
  title     = {Genetic, individual, and familial risk correlates of brain network controllability in major depressive disorder},
  author    = {Hahn, Tim and Winter, Nils R and Ernsting, Jan and Gruber, Marius and Mauritz, Marco J and Fisch, Lukas and Leenings, Ramona and Sarink, Kelvin and Blanke, Julian and Holstein, Vincent and Emden, Daniel and Beisemann, Marie and Opel, Nils and Grotegerd, Dominik and Meinert, Susanne and Heindel, Walter and Witt, Stephanie and Rietschel, Marcella and N{\"o}then, Markus M and Forstner, Andreas J and Kircher, Tilo and Nenadic, Igor and Jansen, Andreas and M{\"u}ller-Myhsok, Bertram and Andlauer, Till F M and Walter, Martin and van den Heuvel, Martijn P and Jamalabadi, Hamidreza and Dannlowski, Udo and Repple, Jonathan},
  journal   = {Molecular Psychiatry},
  volume    = {28},
  number    = {3},
  pages     = {1057--1063},
  year      = {2023},
  publisher = {Nature Publishing Group},
  doi       = {10.1038/s41380-022-01936-6}
}

@article{wang2022alterations,
  title     = {Alterations in white matter network dynamics in patients with schizophrenia and bipolar disorder},
  author    = {Wang, Bin and Zhang, Shanshan and Yu, Xuexue and Niu, Yan and Niu, Jinliang and Li, Dandan and Zhang, Shan and Xiang, Jie and Yan, Ting and Yang, Jiajia and Wu, Jinglong and Liu, Miaomiao},
  journal   = {Human Brain Mapping},
  volume    = {43},
  number    = {13},
  pages     = {3909--3922},
  year      = {2022},
  publisher = {Wiley Online Library},
  doi       = {10.1002/hbm.25892}
}

@article{tang2023altered,
  title     = {Altered controllability of white matter networks and related brain function changes in first-episode drug-naive schizophrenia},
  author    = {Tang, Biqiu and Zhang, Wenjing and Liu, Jiang and Deng, Shikuang and Hu, Na and Li, Siyi and Zhao, Youjin and Liu, Nian and Zeng, Jiaxin and Cao, Hengyi and Sweeney, John A and Gong, Qiyong and Gu, Shi and Lui, Su},
  journal   = {Cerebral Cortex},
  volume    = {33},
  number    = {4},
  pages     = {1527--1535},
  year      = {2023},
  publisher = {Oxford University Press},
  doi       = {10.1093/cercor/bhac421}
}

@article{broeders2024energy,
  title     = {Energy Associated With Dynamic Network Changes in Patients With Multiple Sclerosis and Cognitive Impairment},
  author    = {Broeders, Tommy A A and van Dam, Maureen and Pontillo, Giuseppe and Rauh, Vasco and Douw, Linda and van der Werf, Ysbrand D and Killestein, Joep and Barkhof, Frederik and Vinkers, Christiaan H and Schoonheim, Menno M},
  journal   = {Neurology},
  volume    = {103},
  number    = {9},
  pages     = {e209952},
  year      = {2024},
  publisher = {AAN Enterprises},
  doi       = {10.1212/WNL.0000000000209952}
}

@article{zoller2021psychosis,
  title     = {Structural control energy of resting-state functional brain states reveals less cost-effective brain dynamics in psychosis vulnerability},
  author    = {Z{\"o}ller, Daniela and Sandini, Corrado and Schaer, Marie and Eliez, Stephan and Bassett, Danielle S and Van De Ville, Dimitri},
  journal   = {Human Brain Mapping},
  volume    = {42},
  number    = {7},
  pages     = {2181--2200},
  year      = {2021},
  publisher = {Wiley Online Library},
  doi       = {10.1002/hbm.25358}
}

@article{palmigiano2017flexible,
  title     = {Flexible information routing by transient synchrony},
  author    = {Palmigiano, Agostina and Geisel, Theo and Wolf, Fred and Battaglia, Demian},
  journal   = {Nature Neuroscience},
  volume    = {20},
  number    = {7},
  pages     = {1014--1022},
  year      = {2017},
  publisher = {Nature Publishing Group},
  doi       = {10.1038/nn.4569}
}

@article{cui2020executive,
  title     = {Optimization of energy state transition trajectory supports the development of executive function during youth},
  author    = {Cui, Zaixu and Stiso, Jennifer and Baum, Graham L and Kim, Jason Z and Roalf, David R and Betzel, Richard F and Gu, Shi and Lu, Zhixin and Xia, Cedric H and He, Xiaosong and Ciric, Rastko and Oathes, Desmond J and Moore, Tyler M and Shinohara, Russell T and Ruparel, Kosha and Davatzikos, Christos and Pasqualetti, Fabio and Gur, Raquel E and Gur, Ruben C and Bassett, Danielle S and Satterthwaite, Theodore D},
  journal   = {eLife},
  volume    = {9},
  pages     = {e53060},
  year      = {2020},
  publisher = {eLife Sciences Publications, Ltd},
  doi       = {10.7554/eLife.53060}
}

@article{jamalabadi2023parenthood,
  title     = {Interrelated effects of age and parenthood on whole-brain controllability: protective effects of parenthood in mothers},
  author    = {Jamalabadi, Hamidreza and Hahn, Tim and Winter, Nils R and Nozari, Erfan and Ernsting, Jan and Meinert, Susanne and Leehr, Elisabeth J and Dohm, Katharina and Bauer, Jochen and Pfarr, Julia-Katharina and Stein, Frederike and Thomas-Odenthal, Florian and Brosch, Katharina and Mauritz, Marco and Gruber, Marius and Repple, Jonathan and Kaufmann, Tobias and Krug, Axel and Nenadi{\'c}, Igor and Kircher, Tilo and Dannlowski, Udo and Derntl, Birgit},
  journal   = {Frontiers in Aging Neuroscience},
  volume    = {15},
  pages     = {1085153},
  year      = {2023},
  publisher = {Frontiers Media SA},
  doi       = {10.3389/fnagi.2023.1085153}
}

@article{singleton2024alcohol,
  title     = {Altered Structural Connectivity and Functional Brain Dynamics in Individuals With Heavy Alcohol Use Elucidated via Network Control Theory},
  author    = {Singleton, S Parker and Velidi, Puneet and Schilling, Louisa and Luppi, Andrea I and Jamison, Keith and Parkes, Linden and Kuceyeski, Amy},
  journal   = {Biological Psychiatry: Cognitive Neuroscience and Neuroimaging},
  volume    = {9},
  number    = {10},
  pages     = {1010--1018},
  year      = {2024},
  publisher = {Elsevier},
  doi       = {10.1016/j.bpsc.2024.05.006}
}

@article{yuan2026ocd,
  title     = {Neurochemically informed network control theory reveals energetic dysregulation of altered brain state dynamics in obsessive-compulsive disorder},
  author    = {Yuan, Dongling and Liao, Haiyan and Zhang, Yi and Wang, Zhiyan and Han, Yan and Zhang, Douyu and Yu, Qianmei and Fan, Jie and Zhu, Xiongzhao},
  journal   = {Psychiatry and Clinical Neurosciences},
  year      = {2026},
  publisher = {Wiley Online Library},
  doi       = {10.1111/pcn.70069}
}

@article{parkes2021psychosis,
  title     = {Network Controllability in Transmodal Cortex Predicts Positive Psychosis Spectrum Symptoms},
  author    = {Parkes, Linden and Moore, Tyler M and Calkins, Monica E and Cieslak, Matthew and Roalf, David R and Wolf, Daniel H and Gur, Ruben C and Gur, Raquel E and Satterthwaite, Theodore D and Bassett, Danielle S},
  journal   = {Biological Psychiatry},
  volume    = {90},
  number    = {6},
  pages     = {409--418},
  year      = {2021},
  publisher = {Elsevier},
  doi       = {10.1016/j.biopsych.2021.03.016}
}

@article{jeganathan2018bipolar,
  title     = {Fronto-limbic dysconnectivity leads to impaired brain network controllability in young people with bipolar disorder and those at high genetic risk},
  author    = {Jeganathan, Jayson and Perry, Alistair and Bassett, Danielle S and Roberts, Gloria and Mitchell, Philip B and Breakspear, Michael},
  journal   = {NeuroImage: Clinical},
  volume    = {19},
  pages     = {71--81},
  year      = {2018},
  publisher = {Elsevier},
  doi       = {10.1016/j.nicl.2018.03.032}
}

@article{niu2026depression,
  title     = {Brain energetic landscapes shape state dysregulation in major depressive disorder: a morphological network controllability perspective},
  author    = {Niu, Jinpeng and Xia, Jie and Liu, Qingjin and He, Yaohui and Li, Wei and Chen, Kangjia and Zhang, Xi and Qiu, Jiang and Chen, Huafu and Li, Jiao and Liao, Wei},
  journal   = {Translational Psychiatry},
  year      = {2026},
  publisher = {Nature Publishing Group},
  doi       = {10.1038/s41398-026-04025-2}
}

@article{fang2022personalizing,
  title     = {Personalizing repetitive transcranial magnetic stimulation for precision depression treatment based on functional brain network controllability and optimal control analysis},
  author    = {Fang, Feng and Godlewska, Beata and Cho, Raymond Y and Savitz, Sean I and Selvaraj, Sudhakar and Zhang, Yingchun},
  journal   = {NeuroImage},
  volume    = {260},
  pages     = {119465},
  year      = {2022},
  publisher = {Elsevier},
  doi       = {10.1016/j.neuroimage.2022.119465}
}

@article{wang2025eeg,
  title     = {Personalized {EEG}-guided brain stimulation targeting in major depression via network controllability and multi-objective optimization},
  author    = {Wang, Aihua and Sun, Jingnan},
  journal   = {BMC Psychiatry},
  volume    = {25},
  number    = {1},
  pages     = {723},
  year      = {2025},
  publisher = {Springer},
  doi       = {10.1186/s12888-025-07171-x}
}

@article{khambhati2019stimulation,
  title     = {Functional control of electrophysiological network architecture using direct neurostimulation in humans},
  author    = {Khambhati, Ankit N and Kahn, Ari E and Costantini, Julia and Ezzyat, Youssef and Solomon, Ethan A and Gross, Robert E and Jobst, Barbara C and Sheth, Sameer A and Zaghloul, Kareem A and Worrell, Gregory and Seger, Sarah and Lega, Bradley C and Weiss, Shennan and Sperling, Michael R and Gorniak, Richard and Das, Sandhitsu R and Stein, Joel M and Rizzuto, Daniel S and Kahana, Michael J and Lucas, Timothy H and Davis, Kathryn A and Tracy, Joseph I and Bassett, Danielle S},
  journal   = {Network Neuroscience},
  volume    = {3},
  number    = {3},
  pages     = {848--877},
  year      = {2019},
  publisher = {MIT Press},
  doi       = {10.1162/netn_a_00089}
}

@article{hahn2023ect,
  title     = {Towards a network control theory of electroconvulsive therapy response},
  author    = {Hahn, Tim and Jamalabadi, Hamidreza and Nozari, Erfan and Winter, Nils R and Ernsting, Jan and Gruber, Marius and Mauritz, Marco J and Grumbach, Pascal and Fisch, Lukas and Leenings, Ramona and Sarink, Kelvin and Blanke, Julian and Vennekate, Leon Kleine and Emden, Daniel and Opel, Nils and Grotegerd, Dominik and Enneking, Verena and Meinert, Susanne and Borgers, Tiana and Klug, Melissa and Leehr, Elisabeth J and Dohm, Katharina and Heindel, Walter and Gross, Joachim and Dannlowski, Udo and Redlich, Ronny and Repple, Jonathan},
  journal   = {PNAS Nexus},
  volume    = {2},
  number    = {2},
  pages     = {pgad032},
  year      = {2023},
  publisher = {Oxford University Press},
  doi       = {10.1093/pnasnexus/pgad032}
}

@article{wasilczuk2024hormonal,
  title     = {Hormonal basis of sex differences in anesthetic sensitivity},
  author    = {Wasilczuk, Andrzej Z and Rinehart, Cole and Aggarwal, Adeeti and Stone, Martha E and Mashour, George A and Avidan, Michael S and Kelz, Max B and Proekt, Alex},
  journal   = {Proceedings of the National Academy of Sciences},
  volume    = {121},
  number    = {3},
  pages     = {e2312913120},
  year      = {2024},
  publisher = {National Academy of Sciences},
  doi       = {10.1073/pnas.2312913120}
}

@article{mesulam1998sensation,
  title     = {From sensation to cognition},
  author    = {Mesulam, M Marsel},
  journal   = {Brain},
  volume    = {121},
  number    = {6},
  pages     = {1013--1052},
  year      = {1998},
  publisher = {Oxford University Press},
  doi       = {10.1093/brain/121.6.1013}
}

@article{margulies2016situating,
  title     = {Situating the default-mode network along a principal gradient of macroscale cortical organization},
  author    = {Margulies, Daniel S and Ghosh, Satrajit S and Goulas, Alexandros and Falkiewicz, Marcel and Huntenburg, Julia M and Langs, Georg and Bezgin, Gleb and Eickhoff, Simon B and Castellanos, F Xavier and Petrides, Michael and Jefferies, Elizabeth and Smallwood, Jonathan},
  journal   = {Proceedings of the National Academy of Sciences},
  volume    = {113},
  number    = {44},
  pages     = {12574--12579},
  year      = {2016},
  publisher = {National Academy of Sciences},
  doi       = {10.1073/pnas.1608282113}
}

@article{mitchell2013gendered,
  title     = {Gendered citation patterns in international relations journals},
  author    = {Mitchell, Sara McLaughlin and Lange, Samantha and Brus, Holly},
  journal   = {International Studies Perspectives},
  volume    = {14},
  number    = {4},
  pages     = {485--492},
  year      = {2013},
  publisher = {Wiley},
  doi       = {10.1111/insp.12026}
}

@article{maliniak2013gender,
  title     = {The gender citation gap in international relations},
  author    = {Maliniak, Daniel and Powers, Ryan and Walter, Barbara F},
  journal   = {International Organization},
  volume    = {67},
  number    = {4},
  pages     = {889--922},
  year      = {2013},
  publisher = {Cambridge University Press},
  doi       = {10.1017/S0020818313000209}
}

@article{caplar2017quantitative,
  title     = {Quantitative evaluation of gender bias in astronomical publications from citation counts},
  author    = {Caplar, Neven and Tacchella, Sandro and Birrer, Simon},
  journal   = {Nature Astronomy},
  volume    = {1},
  number    = {6},
  pages     = {0141},
  year      = {2017},
  publisher = {Nature Publishing Group},
  doi       = {10.1038/s41550-017-0141}
}

@article{dion2018gendered,
  title     = {Gendered citation patterns across political science and social science methodology fields},
  author    = {Dion, Michelle L and Sumner, Jane Lawrence and Mitchell, Sara McLaughlin},
  journal   = {Political Analysis},
  volume    = {26},
  number    = {3},
  pages     = {312--327},
  year      = {2018},
  publisher = {Cambridge University Press},
  doi       = {10.1017/pan.2018.12}
}

@article{dworkin2020extent,
  title     = {The extent and drivers of gender imbalance in neuroscience reference lists},
  author    = {Dworkin, Jordan D and Linn, Kristin A and Teich, Erin G and Zurn, Perry and Shinohara, Russell T and Bassett, Danielle S},
  journal   = {Nature Neuroscience},
  volume    = {23},
  number    = {8},
  pages     = {918--926},
  year      = {2020},
  publisher = {Nature Publishing Group},
  doi       = {10.1038/s41593-020-0658-y}
}

@article{bertolero2021racial,
  title     = {Racial and ethnic imbalance in neuroscience reference lists and intersections with gender},
  author    = {Bertolero, Maxwell A and Dworkin, Jordan D and David, Sophia U and L{\'o}pez Lloreda, Claudia and Srivastava, Pragya and Stiso, Jennifer and Zhou, Dale and Dzirasa, Kafui and Fair, Damien A and Kaczkurkin, Antonia N and Marlin, Bianca Jones and Shohamy, Daphna and Uddin, Lucina Q and Zurn, Perry and Bassett, Danielle S},
  journal   = {bioRxiv},
  year      = {2020},
  publisher = {Cold Spring Harbor Laboratory},
  doi       = {10.1101/2020.10.12.336230}
}

@article{wang2021gendered,
  title     = {Gendered citation practices in the field of communication},
  author    = {Wang, Xinyi and Dworkin, Jordan D and Zhou, Dale and Stiso, Jennifer and Falk, Emily B and Bassett, Danielle S and Zurn, Perry and Lydon-Staley, David M},
  journal   = {Annals of the International Communication Association},
  volume    = {45},
  number    = {2},
  pages     = {134--153},
  year      = {2021},
  publisher = {Taylor \& Francis},
  doi       = {10.1080/23808985.2021.1960180}
}

@article{chatterjee2021gender,
  title     = {Gender disparity in citations in high-impact journal articles},
  author    = {Chatterjee, Paula and Werner, Rachel M},
  journal   = {JAMA Network Open},
  volume    = {4},
  number    = {7},
  pages     = {e2114509},
  year      = {2021},
  publisher = {American Medical Association},
  doi       = {10.1001/jamanetworkopen.2021.14509}
}

@article{fulvio2021imbalance,
  title     = {Gender (im)balance in citation practices in cognitive neuroscience},
  author    = {Fulvio, Jacqueline M and Akinnola, Ileri and Postle, Bradley R},
  journal   = {Journal of Cognitive Neuroscience},
  volume    = {33},
  number    = {1},
  pages     = {3--7},
  year      = {2021},
  publisher = {MIT Press},
  doi       = {10.1162/jocn_a_01643}
}

@misc{zhou2020gender,
  title        = {Gender Diversity Statement and Code Notebook v1.0},
  author       = {Zhou, Dale and Cornblath, Eli J and Stiso, Jennifer and Teich, Erin G and Dworkin, Jordan D and Blevins, Ann S and Bassett, Danielle S},
  year         = {2020},
  howpublished = {Zenodo},
  doi          = {10.5281/zenodo.3672110}
}

@inproceedings{ambekar2009name,
  title     = {Name-ethnicity classification from open sources},
  author    = {Ambekar, Anurag and Ward, Charles and Mohammed, Jahangir and Male, Swapna and Skiena, Steven},
  booktitle = {Proceedings of the 15th ACM SIGKDD International Conference on Knowledge Discovery and Data Mining},
  pages     = {49--58},
  year      = {2009},
  publisher = {Association for Computing Machinery},
  doi       = {10.1145/1557019.1557032}
}

@article{sood2018predicting,
  title     = {Predicting race and ethnicity from the sequence of characters in a name},
  author    = {Sood, Gaurav and Laohaprapanon, Suriyan},
  journal   = {arXiv preprint arXiv:1805.02109},
  year      = {2018},
  doi       = {10.48550/arXiv.1805.02109}
}

\pagebreak

\section*{Acknowledgments}
Data were provided in part by the Human Connectome Project, WU-Minn Consortium (Principal Investigators: David Van Essen and Kamil Ugurbil; 1U54MH091657) funded by the 16 National Institutes of Health (NIH) Institutes and Centers that support the NIH Blueprint for Neuroscience Research; and by the McDonnell Center for Systems Neuroscience at Washington University. S.M. and D.S.B. acknowledge support from the Army Research Office MURI program (through grant number W911NF2410228).

\section*{Contributions}
Conceptualization: S.M., C.G.A., D.S.B. Methodology: S.M., C.G.A., P.F., D.S.B. Data Analyses, Writing – Original Draft: S.M., C.G.A., P.F. Writing – Review \& Editing: S.M., C.G.A., P.F., M.K., J.D.,F.P., D.S.B. Supervision: D.S.B.

\section*{Competing interests}
The authors declare no competing interests.

\end{document}


\maketitle
\pagebreak

\section*{Supplementary Figures}

\begin{figure}[htb]
    \centering
    \includegraphics[width=1\textwidth]{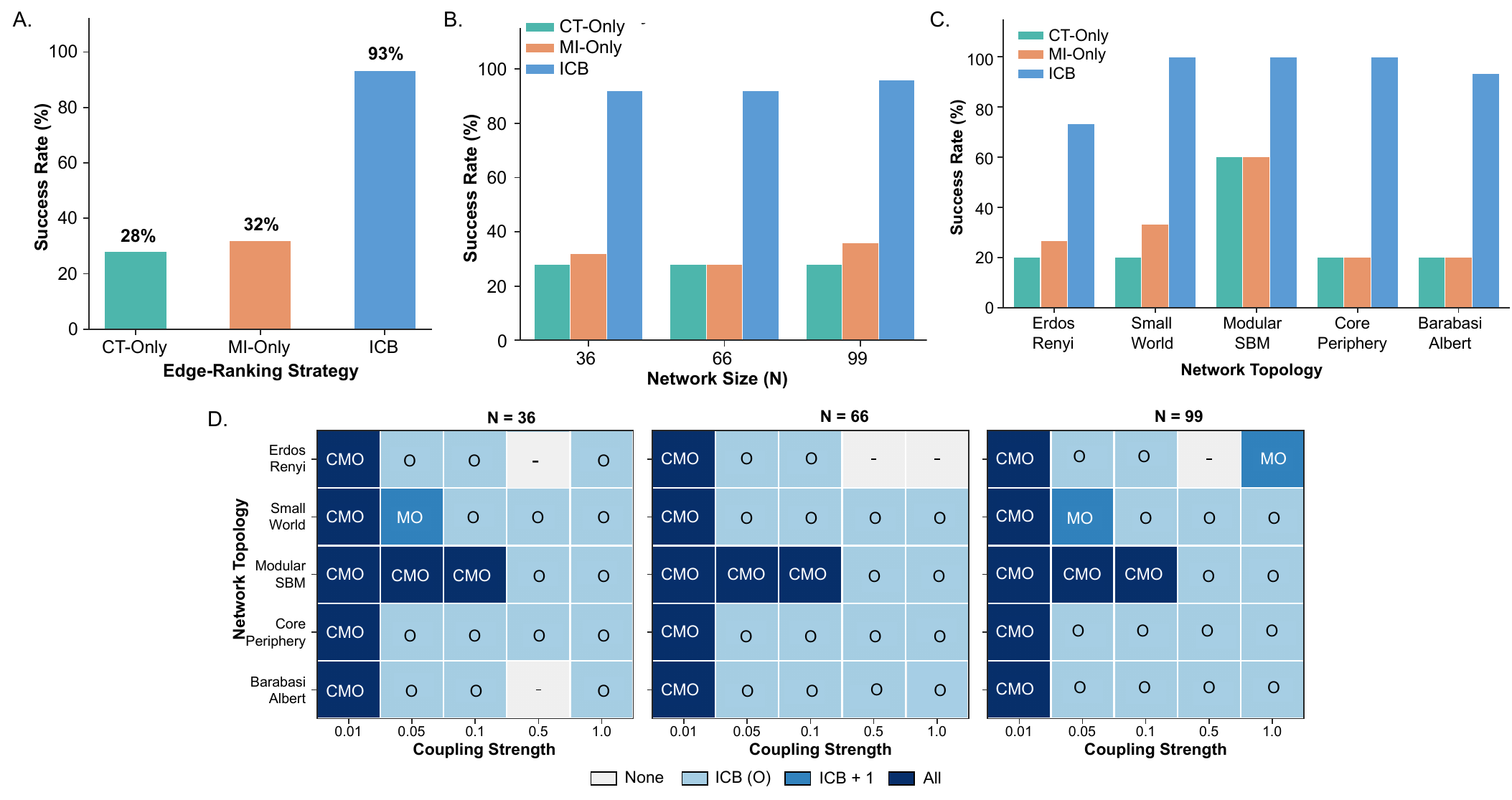}
    \caption{\textbf{Ablation analysis confirms that both information-theoretic and controllability components are necessary for QUIET performance.}
    \textbf{A.} Overall success rate (percentage of topology $\times$ scale $\times$ coupling-strength configurations achieving significant synchronization, $p<0.01$).
    \textbf{B.} Success rate stratified by network size ($N$ = 36, 66, 99). QUIET maintains high performance across all scales, whereas single-component rankings remain below 40\%.
    \textbf{C.} Success rate stratified by network topology.
    \textbf{D.} Configuration-level success maps for each topology $\times$ coupling-strength combination at $N$ = 36, 66, and 99. Controllability only (C), Mutual Information only (M), and the full QUIET ranking (O).}
    \label{SuppFig1}
\end{figure}

\begin{figure}[htb]
    \centering
    \includegraphics[width=1\textwidth]{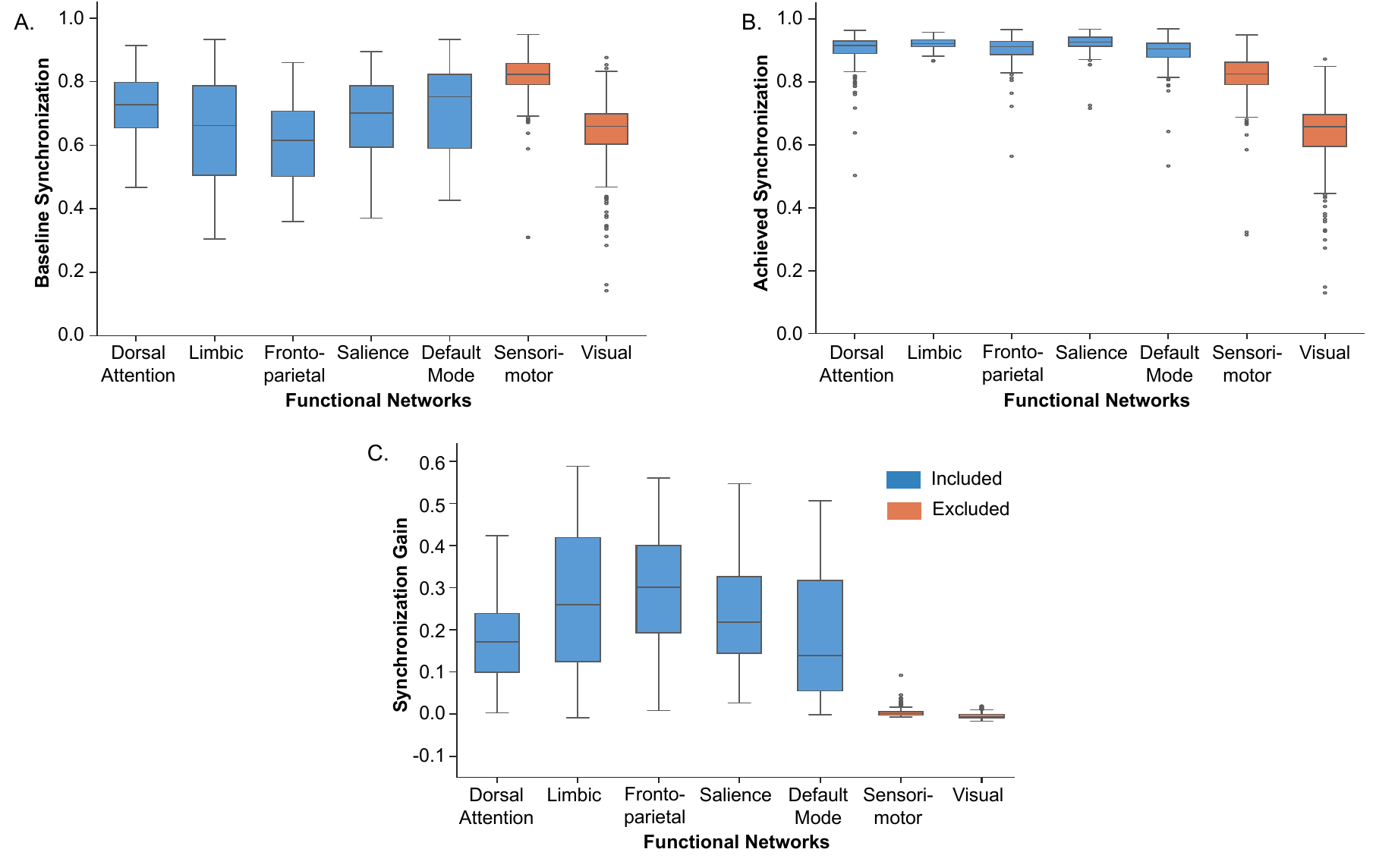}
    \caption{\textbf{Ceiling and floor effects in sensorimotor and visual networks, respectively.}
    \textbf{A.} Baseline synchronization (phase-locking value, PLV) across all seven networks. The sensorimotor network exhibits the highest median baseline PLV, indicating a ceiling effect.
    \textbf{B.} Optimised synchronization after QUIET-driven perturbation.
    \textbf{C.} Synchronization gain (achieved $-$ baseline PLV). Both the sensorimotor and visual networks show near-zero gain, confirming ceiling and floor effects, respectively. Orange boxes denote excluded networks; blue boxes denote included networks.}
    \label{SuppFig2}
\end{figure}

\begin{figure}[htb]
    \centering
    \includegraphics[width=1\textwidth]{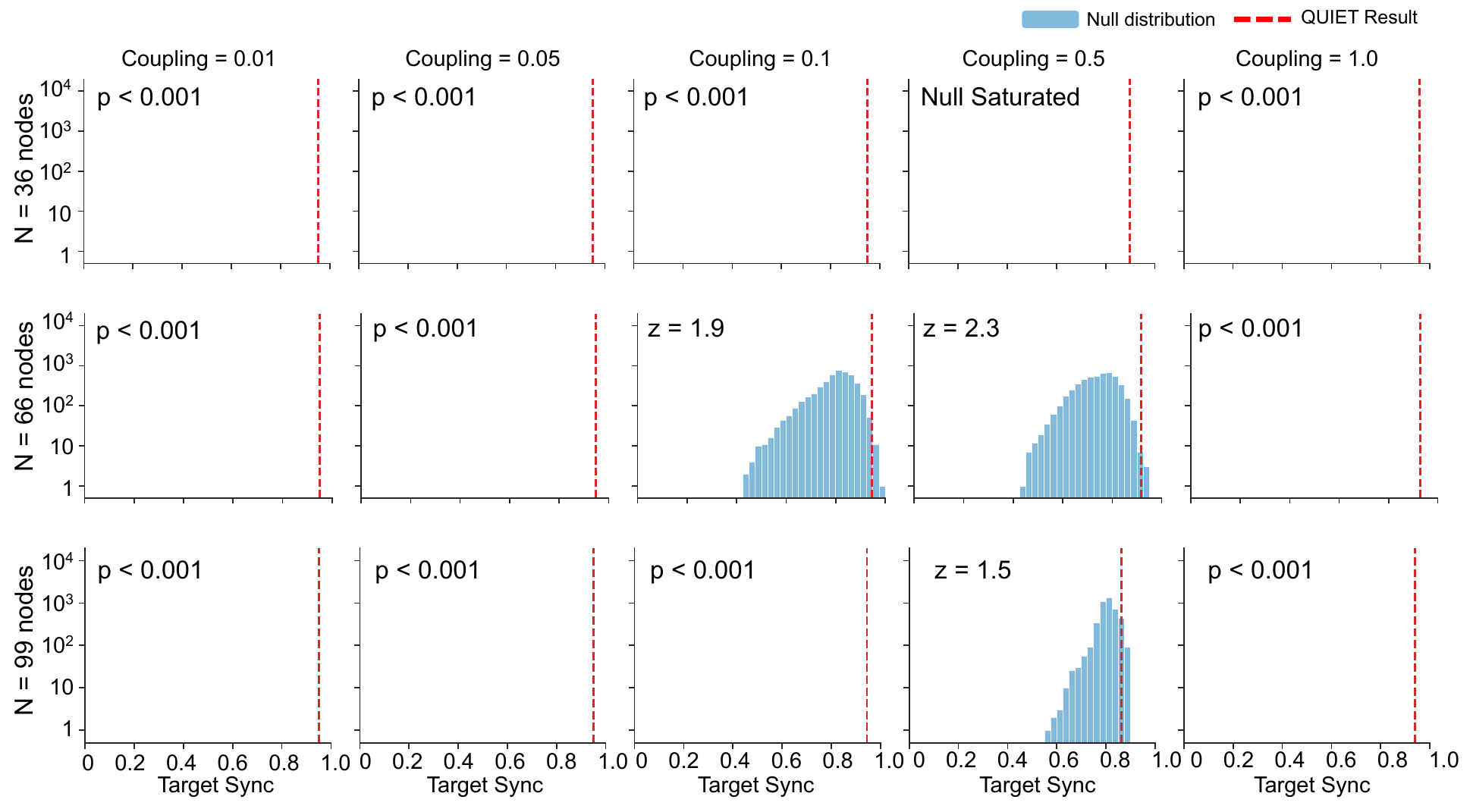}
    \caption{\textbf{Null-model distributions of target synchronization for the Erd\H{o}s-R\'{e}nyi topology.}
    Each panel shows the distribution of target-module phase-locking values (PLV) obtained from 10,000 random edge-set permutations (blue histograms; counts on $\log_{10}$ scale) for an Erd\H{o}s-R\'{e}nyi network at a given size (rows: $N$ = 36, 66, 99 nodes) and coupling strength (columns: $g$ = 0.01, 0.05, 0.1, 0.5, 1.0). The red dashed line marks the PLV achieved by the QUIET-ranked edge set. To enable a fair comparison, the null distribution is restricted to permutations whose rest mean absolute error (MAE) is at or below QUIET's rest MAE. Panel annotations reflect three statistical states: $z = X.X$ when the comparison yields a meaningful test statistic; $p < 0.001$ when QUIET exceeds the null at the strictest joint sync-MAE criterion (the histogram is suppressed for visual clarity in these highly significant configurations); and null saturated when all random selections produce essentially identical synchronization, rendering the QUIET-vs-null comparison non-informative. Cells marked null saturated indicate that no random permutation survived the joint MAE filter, not that QUIET failed the null test. The Erd\H{o}s-R\'{e}nyi topology shows distinct null distributions at the lowest coupling strengths and a single saturated configuration ($N$ = 36, $g$ = 0.5).}
    \label{SuppFig3}
\end{figure}

\begin{figure}[htb]
    \centering
    \includegraphics[width=1\textwidth]{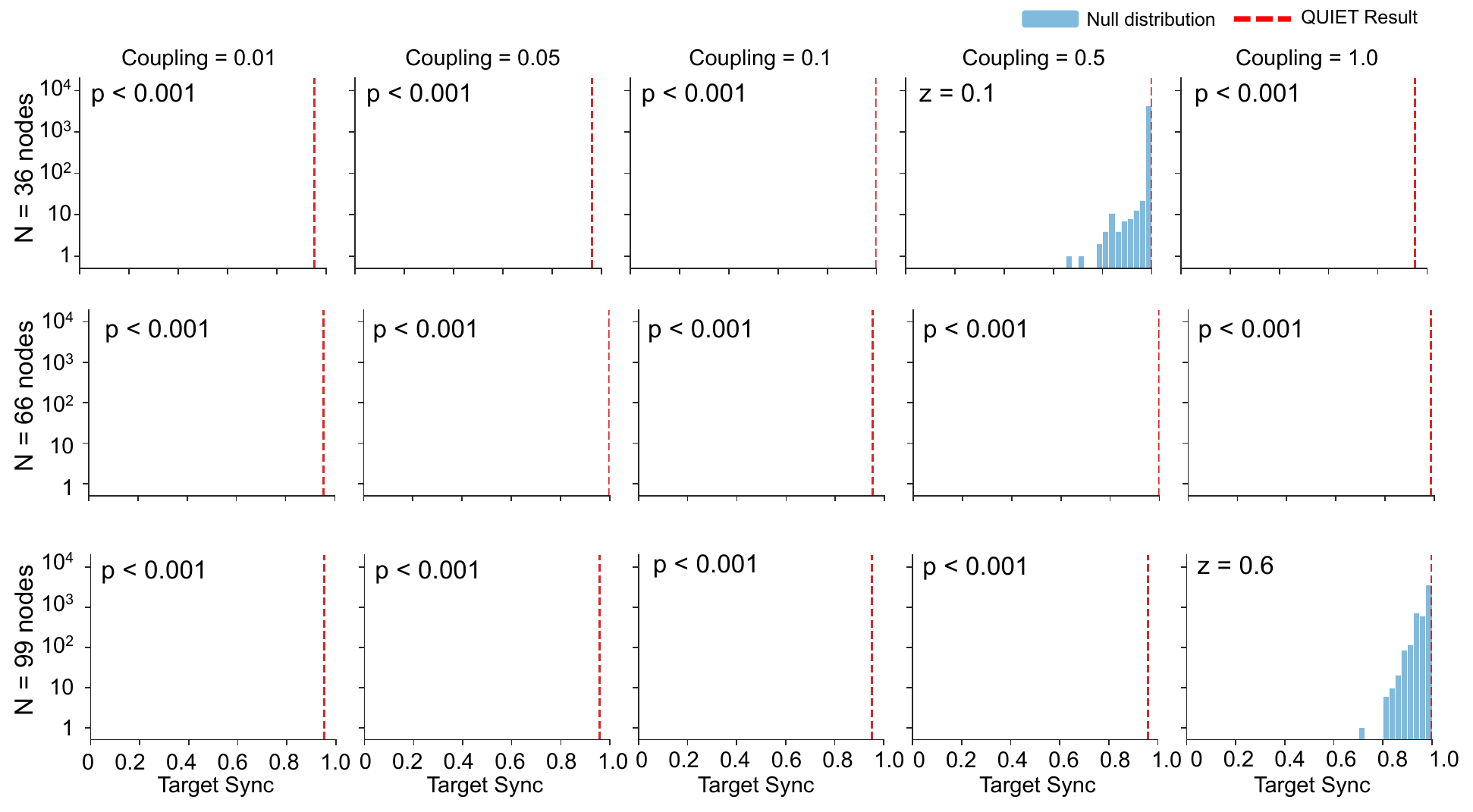}
    \caption{\textbf{Null-model distributions of target synchronization for the small-world topology.}
    Each panel shows the distribution of target-module phase-locking values (PLV) obtained from 10,000 random edge-set permutations (blue histograms; counts on $\log_{10}$ scale) for a Watts-Strogatz small-world network at a given size (rows: $N$ = 36, 66, 99 nodes) and coupling strength (columns: $g$ = 0.01, 0.05, 0.1, 0.5, 1.0). The red dashed line marks the PLV achieved by the QUIET-ranked edge set, and the null distribution is restricted to permutations whose rest MAE is at or below QUIET's rest MAE. Panel annotations follow the convention introduced in Supplementary Fig.~\ref{SuppFig3}: $z = X.X$ for meaningful test statistics, $p < 0.001$ for highly significant configurations (histogram suppressed), and null saturated for degenerate nulls. The small-world topology yields predominantly highly significant out performance across $N$ and $g$, with two configurations ($N$ = 36, $g$ = 0.5; $N$ = 99, $g$ = 1.0) where the conditional null distribution overlaps QUIET sufficiently to warrant explicit display.}
    \label{SuppFig4}
\end{figure}

\begin{figure}[htb]
    \centering
    \includegraphics[width=1\textwidth]{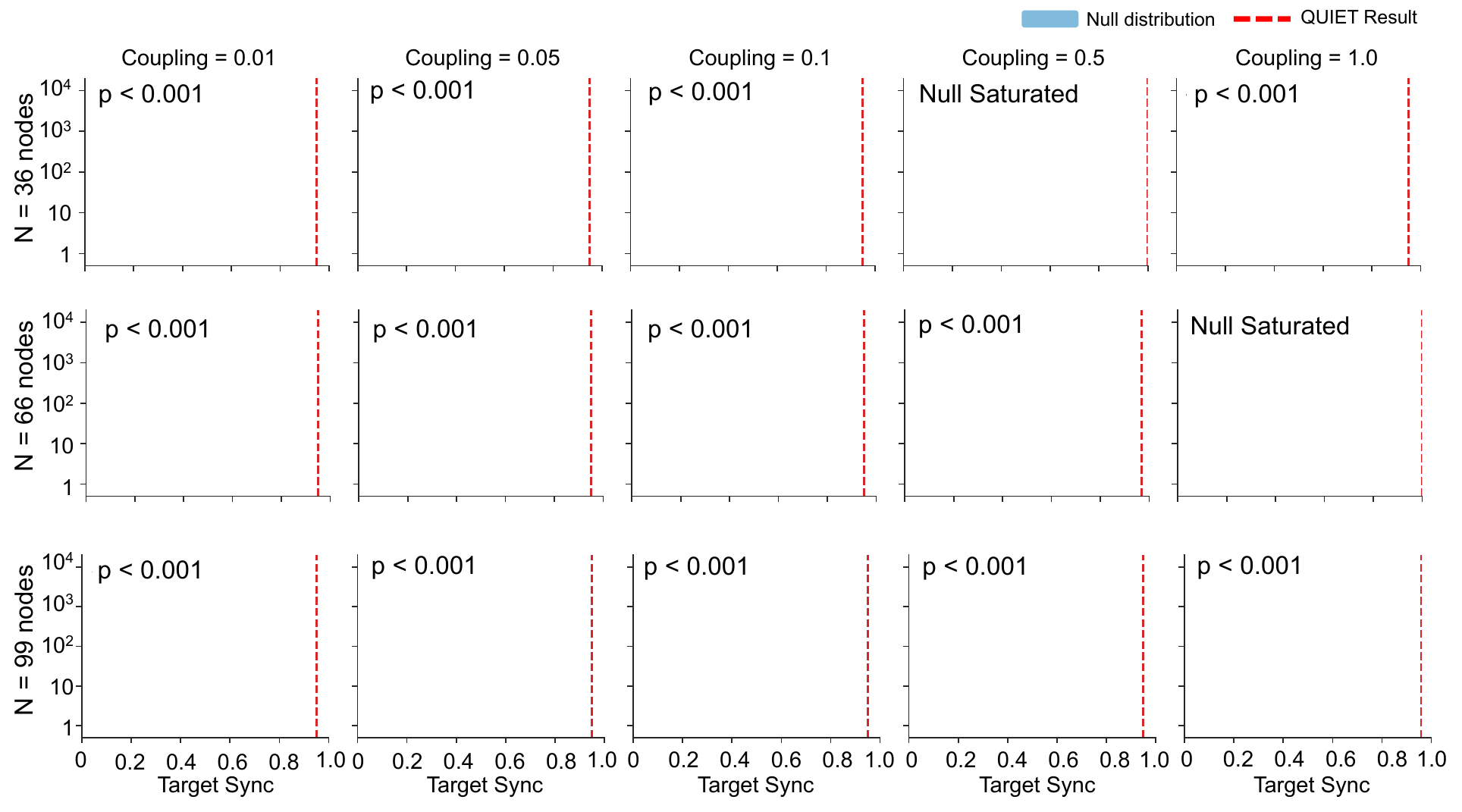}
    \caption{\textbf{Null-model distributions of target synchronization for the modular stochastic block model topology.}
    Each panel shows the distribution of target-module phase-locking values (PLV) obtained from 10,000 random edge-set permutations (blue histograms; counts on $\log_{10}$ scale) for a modular stochastic block model network at a given size (rows: $N$ = 36, 66, 99 nodes) and coupling strength (columns: $g$ = 0.01, 0.05, 0.1, 0.5, 1.0). The red dashed line marks the PLV achieved by the QUIET-ranked edge set, and the null distribution is restricted to permutations whose rest MAE is at or below QUIET's rest MAE. Panel annotations follow the convention introduced in Supplementary Fig.~\ref{SuppFig3}: $z = X.X$ for meaningful test statistics, $p < 0.001$ for highly significant configurations (histogram suppressed), and ``null saturated'' for degenerate nulls. Modular networks exhibit clearly bounded null distributions at low coupling, and QUIET consistently exceeds the conditional null mean across both $N$ and $g$.}
    \label{SuppFig5}
\end{figure}

\begin{figure}[htb]
    \centering
    \includegraphics[width=1\textwidth]{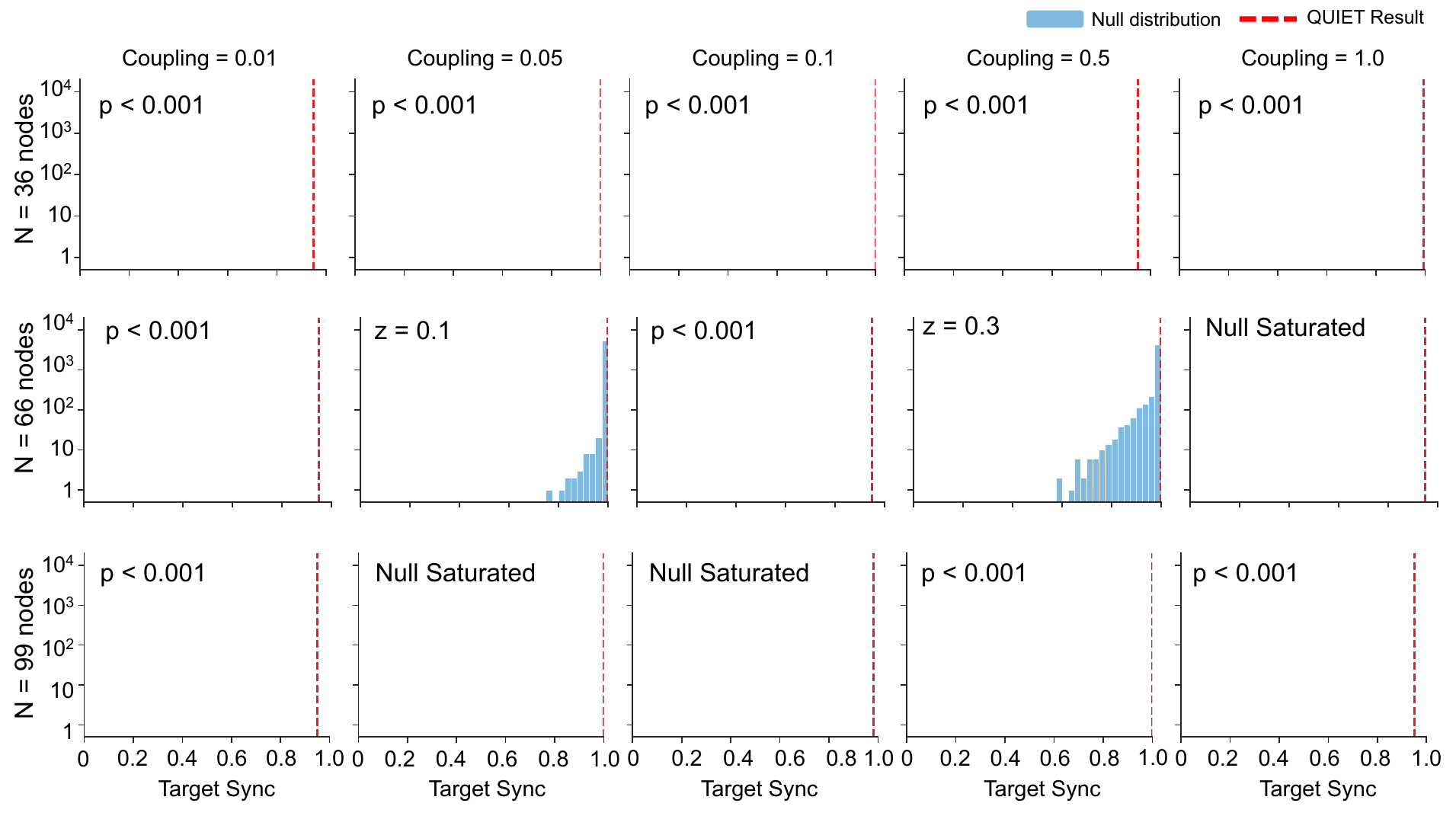}
    \caption{\textbf{Null-model distributions of target synchronization for the core-periphery topology.}
    Each panel shows the distribution of target-module phase-locking values (PLV) obtained from 10,000 random edge-set permutations (blue histograms; counts on $\log_{10}$ scale) for a core-periphery network at a given size (rows: $N$ = 36, 66, 99 nodes) and coupling strength (columns: $g$ = 0.01, 0.05, 0.1, 0.5, 1.0). The red dashed line marks the PLV achieved by the QUIET-ranked edge set, and the null distribution is restricted to permutations whose rest MAE is at or below QUIET's rest MAE. Panel annotations follow the convention introduced in Supplementary Fig.~\ref{SuppFig3}: $z = X.X$ for meaningful test statistics, $p < 0.001$ for highly significant configurations (histogram suppressed), and null saturated for degenerate nulls. The core-periphery topology shows that QUIET-ranked edges achieve highly significant out performance across most configurations, with null distributions concentrating near the upper PLV range as coupling strength grows.}
    \label{SuppFig6}
\end{figure}

\begin{figure}[htb]
    \centering
    \includegraphics[width=1\textwidth]{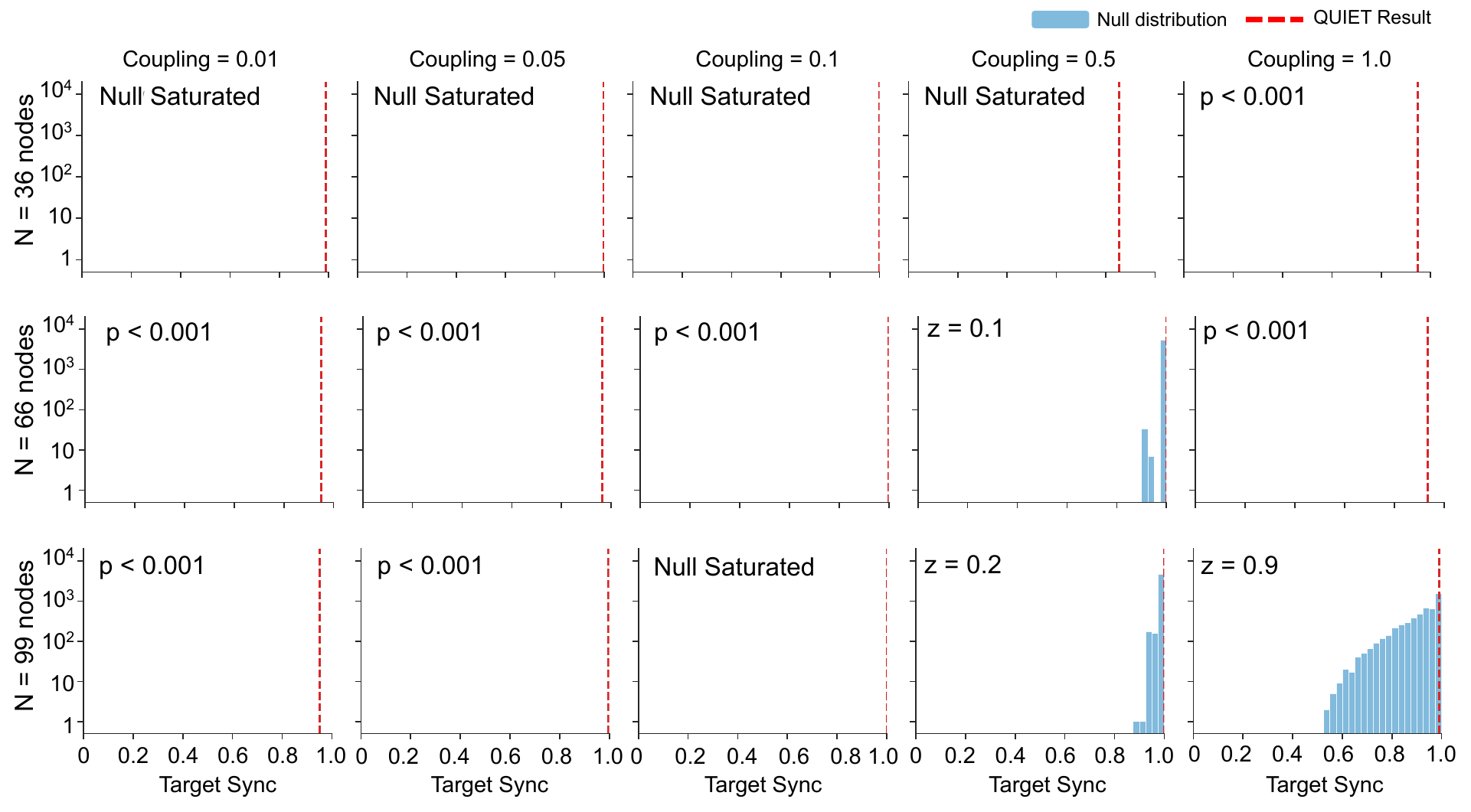}
    \caption{\textbf{Null-model distributions of target synchronization for the Barab\'{a}si-Albert topology.}
    Each panel shows the distribution of target-module phase-locking values (PLV) obtained from 10,000 random edge-set permutations (blue histograms; counts on $\log_{10}$ scale) for a Barab\'{a}si-Albert scale-free network at a given size (rows: $N$ = 36, 66, 99 nodes) and coupling strength (columns: $g$ = 0.01, 0.05, 0.1, 0.5, 1.0). The red dashed line marks the PLV achieved by the QUIET-ranked edge set, and the null distribution is restricted to permutations whose rest MAE is at or below QUIET's rest MAE. Panel annotations follow the convention introduced in Supplementary Fig.~\ref{SuppFig3}: $z = X.X$ for meaningful test statistics, $p < 0.001$ for highly significant configurations (histogram suppressed), and null saturated for degenerate nulls. Scale-free networks exhibit a higher fraction of saturated null configurations than the other topologies, consistent with the known difficulty of controlling heterogeneous degree distributions in which a small number of hubs dominate the dynamics.}
    \label{SuppFig7}
\end{figure}

\begin{figure}[htb]
    \centering
    \includegraphics[width=1\textwidth]{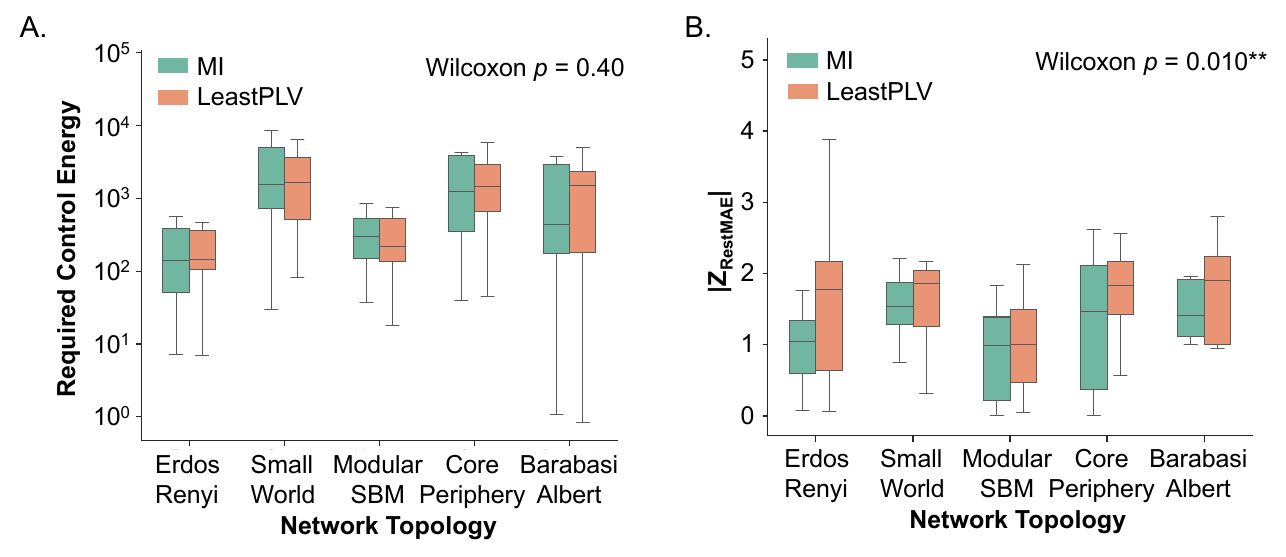}
    \caption{\textbf{The QUIET framework's set-level outcomes are robust to the choice of functional channel, and mutual information yields significantly better off-target stability than a baseline pairwise PLV alternative.}
    \textbf{A.} Required control energy ($E^*$, $\log_{10}$ scale) across the 75-configuration synthetic grid stratified by network topology and functional channel. Box plots show median, interquartile range, and $1.5\times$ IQR whiskers, with outlier markers suppressed. The two pipelines produce statistically indistinguishable control energies across the grid (paired Wilcoxon signed-rank $p = 0.40$, 36/36/0 win/loss/tie).
    \textbf{B.} Absolute $z$-score of rest-cluster PLV deviation from baseline ($|z_{\mathrm{rest\,MAE}}|$). Lower values indicate more selective perturbations that leave off-target dynamics undisturbed. Mutual information achieves significantly lower rest-cluster disturbance than LeastPLV across the grid (paired Wilcoxon $p = 0.010$, mean $|z_{\mathrm{rest\,MAE}}| = 1.28$ for MI vs $1.51$ for LeastPLV).}
    \label{SuppFig8}
\end{figure}

\clearpage
\section*{Supplementary Tables}

\begin{table}[htb]
\centering
\caption{\textbf{Behavioral measures used in the energy-behavior correlation analysis.} Display names, corresponding HCP score names, and the functional network exhibiting the strongest correlation with QUIET-derived control energy.}
\label{SuppTab1}
\begin{tabular}{c c c c c}
\hline
\textbf{Display Name} & \textbf{Score Name} & \textbf{Network} & \textbf{$\rho$} & \textbf{$p$} \\
\hline
Fluid Intelligence       & CogFluidComp\_AgeAdj    & Salience        & $-0.25$ & $0.014$ \\
Cognitive Flexibility    & CardSort\_AgeAdj        & Salience        & $-0.27$ & $0.008$ \\
Inhibitory Control       & Flanker\_AgeAdj         & Salience        & $-0.14$ & $0.166$ \\
Processing Speed         & ProcSpeed\_AgeAdj       & Salience        & $-0.19$ & $0.053$ \\
Patience                 & DDisc\_AUC\_200\_AgeAdj  & Default Mode    & $-0.27$ & $0.008$ \\
Episodic Memory          & PicSeq\_AgeAdj          & Frontoparietal  & $-0.23$ & $0.024$ \\
Perceived Rejection      & PercReject\_AgeAdj      & Limbic          & $+0.19$ & $0.054$ \\
Aggression               & AngAggr\_AgeAdj         & Salience        & $+0.21$ & $0.033$ \\
Perceived Stress         & PercStress\_AgeAdj      & Default Mode    & $+0.26$ & $0.009$ \\
Loneliness               & Loneliness\_AgeAdj      & Limbic          & $+0.15$ & $0.131$ \\
\hline
\end{tabular}
\end{table}

\bibliography{ref}